\documentclass[prb,preprint, longbibliography,superscriptaddress]{revtex4-1}
\usepackage{amsmath}  
\usepackage{amsfonts} 
\usepackage{graphicx} 

\usepackage[utf8]{inputenc} \usepackage{amsmath} \usepackage{bm} \usepackage{tikz}
\usepackage{pgfplots}

\newcommand{\nc}{\protect{\nonumber}\\}
\newcommand{\cd}[1]{c_{#1}^{\dagger}} \newcommand{\cn}[1]{c_{#1}^{\phantom{\dagger}}}

\newcommand{\ad}[1]{a_{#1}^{\dagger}} \newcommand{\an}[1]{a_{#1}^{\phantom{\dagger}}}
\newcommand{\bd}[1]{b_{#1}^{\dagger}} \newcommand{\bn}[1]{b_{#1}^{\phantom{\dagger}}}
\newcommand{\cdu}[1]{c_{#1\uparrow}^{\dagger}}

\newcommand{\cdd}[1]{c_{#1\downarrow}^{\dagger}}

\newcommand{\ddg}[1]{d_{#1}^{\dagger}} 
\newcommand{\dn}[1]{d_{#1}^{\phantom{\dagger}}}
  \newcommand{\hnocc}[1]{\mathbf{n_{#1}^{\phantom{\dagger}}}}

\newcommand{\hbnocc}[1]{\mathbf{\bar n_{#1}^{\phantom{\dagger}}}}

\newcommand{\ket}[1]{\vert
#1\rangle}

 \newcommand{\kvac}{\vert
{\mbox{\o}}\rangle} \newcommand{\bvac}{\langle \mbox{\o}\vert}
\newcommand{\sub}[1]{_{#1}^{\phantom{\dagger}}}
\newcommand{\super}[1]{^{#1}_{\phantom{\dagger}}} \newcommand{\ups}{\uparrow}
\newcommand{\dos}{\downarrow} \newcommand{\hc}{\mbox{H.~c.}}  \newcommand{\rhs}{right-hand
side} \newcommand{\lhs}{left-hand side} \newcommand{\tun}{t} \newcommand{\ttbc}{\tau}
\newcommand{\ttbcs}{\tau^*} \newcommand{\tch}{\tun\sub{0}}
\newcommand{\tchs}{\tun^{*}_{0}} 
 \newcommand{\dd}{\,\mathrm{d}}
 \newcommand{\kocc}{k=\mbox{\footnotesize occ}} 
\newcommand{\pbc}{PBC} \newcommand{\obc}{OBC} \newcommand{\tbc}{TBC}

\begin{document} 
\title{Symmetries and boundary conditions with a twist} \author{Krissia Zawadzki}
\affiliation{Departamento de Física e Ciência Interdisciplinar, Instituto de Física de São Carlos,
  University of São Paulo, Caixa Postal 369, 13560-970 São Carlos, SP, Brazil} \author{Irene
  D'Amico} \affiliation{Department of Physics, University of York, York, YO10\,5DD, United Kingdom}
\affiliation{Departamento de Física e Ciência Interdisciplinar, Instituto de Física de São Carlos,
  University of São Paulo, Caixa Postal 369, 13560-970 São Carlos, SP, Brazil}\author{Luiz
  N. Oliveira} \affiliation{Departamento de Física e Ciência Interdisciplinar, Instituto de Física
  de São Carlos, University of São Paulo, Caixa Postal 369, 13560-970 São Carlos, SP, Brazil}
\date{\today}
\begin{abstract}
  Interest in finite-size systems has risen in the last decades, due to the focus on
  nanotechnological applications and because they are convenient for numerical treatment that can
  subsequently be extrapolated to infinite lattices. Independently of the envisioned application,
  special attention must be given to boundary condition, which may or may not preserve the symmetry
  of the infinite lattice. Here we present a detailed study of the compatibility between boundary
  conditions and conservation laws. The conflict between open boundary conditions and momentum
  conservation is well understood, but we examine other symmetries, as well: we discuss gauge
  invariance, inversion, spin, and particle-hole symmetry and their compatibility with open,
  periodic, and twisted boundary conditions. In the interest of clarity, we develop the reasoning in
  the framework of the one-dimensional half-filled Hubbard model, whose Hamiltonian displays a
  variety of symmetries. Our discussion includes analytical and numerical results. Our analytical
  survey shows that, as a rule, boundary conditions break one or more symmetries of the
  infinite-lattice Hamiltonian. The exception is twisted boundary condition with the special torsion
  $\Theta=\pi L/2$, where $L$ is the lattice size. Our numerical results for the ground-state energy
  at half-filling and the energy gap for $L=2$--$7$ show how the breaking of symmetry affects the
  convergence to the $L\to\infty$ limit. We compare the computed energies and gaps with the exact
  results for the infinite lattice drawn from the Bethe-Ansatz solution. The deviations are
  boundary-condition dependent. The special torsion yields more rapid convergence than open or
  periodic boundary conditions. For sizes as small as $L=7$, the numerical results for twisted
  condition are very close to the $L\to\infty$ limit. We also discuss the ground-state electronic
  density and magnetization at half filling under the three boundary conditions.
\end{abstract}
\maketitle

\section{Introduction}
\label{sec:1} Boundary conditions are of crucial importance to solve physical problems, as they
affect the symmetries of the system and hence may modify fundamental properties, such as ground
state energies and conserved quantities.  For small systems the effect of boundary conditions -- and
of related symmetries -- is particularly acute: this is becoming of more and more practical
relevance as the size of samples considered in experiments is shrinking to the nanoscale, and even
down to just few atoms or spins, spurred by interest in nano and quantum technologies.

In this respect, the importance of the Hubbard model has grown with time. Originally seen as a
sketchy depiction of a strongly correlated solid, the model has found recent experimental expression
e.g. in Bose-Einstein condensates\cite{2016Cheuk1260,2016Boll1257,2016Parsons1253} or ultracold
fermionic atoms.\cite{2015Murmann080402} The (infinite) model exhibits various symmetries. The
Hubbard Hamiltonian conserves charge and spin. In one dimension, it remains invariant under
left-right inversion and therefore conserves parity. The infinite system is invariant under lattice
translations and hence conserves momentum. Finally, if the chemical potential is chosen to make the
number of electrons equal to the number of sites, the Hamiltonian remains invariant under
particle-hole transformation.

Most of the research on the one-dimensional Hubbard Hamiltonian has been focused on the
infinite system. Here, we consider small Hubbard lattices, to compare the effects of different
boundary conditions. Small lattices are in fact important for comparisons to experiments with
Bose-Einstein condensates, molecules, and other physical
systems.\cite{2007Ghirri214405,2010Candini037203,2016Johnson240402,2016Gallemi063626,2016Johnson240402,2016Salfi11342,2016Ferrando-Soria11377}
More specifically, we compute the ground-state energy, energy gap, and electronic and
magnetization densities at half filling for open (\obc), periodic (\pbc), and twisted
(\tbc) boundary conditions for lattices with $L$ ($L=2,3,\ldots,7$) sites. We compare the
results with those determined by the Bethe-Ansatz solution.  Our results show that \tbc\
ensures the fastest convergence to the $L\to\infty$ limit, giving accurate results in most interaction regimes already for chains of only 5 sites. We expect this finding to have
practical value for future numerical treatment of model Hamiltonians. It may also help
identifying under which conditions  a Bose-Einstein condensate or other nanoscale structure
can be used to simulate an infinite Hubbard-model chain.

\section{Overview of twisted boundary conditions}
Twisted boundary conditions are less used and known than open or periodic ones; however,
 we will demonstrate that they are of particular importance for short Hubbard chains.
In this section we summarize their history and usage so far.

In the early 1960's, Kohn found inspiration in the by-then famous paper by Aharonov and Bohm \cite{1959Aharonov485}
and added a magnetic flux threading the center of a ring-shaped system to study its
transport properties.\cite{1964Kohn171} From this formulation he derived a criterion
allowing detection of metal-insulator transitions, of Mott transitions in particular. He
also pointed out that the magnetic flux is equivalent to substituting twisted boundary
condition for the periodic condition defining the ring.

Various analytical developments have directly benefited from Kohn's
formulation.\cite{1965Tho893,1990Shastry243,1990Sutherland1833,1991MaF,1997Shiroishi2288}
More recently, however, numerical applications have given especial prominence to twisted
boundary condition. A method to compute excitation properties of dilute magnetic alloys
was reported three decades ago.\cite{1986Frota7871,YWO90:9403} A few years later, a
procedure applying twisted boundary conditions to the quantum Monte Carlo method
\cite{1993TinkaGammel4437,1996Gros6865,2001Ceperley016702} and allowing efficient,
accurate scaling to the thermodynamical limit of physical properties computed on
relatively small lattices opened new avenues exploited by recent applications in Condensed
Matter,\cite{2008Chiesa086401,2015Santos023632} Nuclear,\cite{2016Schuetrumpf054304} and
High-Energy
Physics.\cite{2004Divitiis408,2005Sachrajda73,2005Bedaque208,2005Bedaque208,2006Flynn313,2007Jiang314,2015Agadjanov118,2015Nitta063,2016Colangelo134}

Twisted boundary condition can be regarded as an extension of Born-von-Karmann, or
periodic, boundary condition. Under periodic boundary condition, opposite ends of a system
are coupled as if they were nearest neighbors inside the system. Under twisted boundary
condition, if the coupling between nearest neighbors is $\tch$, the coupling between the
ends is $\tch\exp(i\Theta)$, where the phase $\Theta$, known as the \emph{torsion}, is a
real number.

\section{One-dimensional Hubbard model}
\label{sec:2} The Hubbard model can be defined on a linear chain, with $L$ sites. Each
site can accommodate up to two electrons. A penalty $U>0$ is imposed on double occupation,
to mimic Coulomb repulsion between electrons of opposite spins, and a coupling $\tch$, a
complex number, allows hopping between a site and its nearest neighbors. The coupling
between the first site ($\ell=1$) and the last one ($\ell=L$) defines the boundary
condition.

\subsection{Hamiltonian}
\label{sec:32}

The model Hamiltonian reads
\begin{align}
   \label{eq:1} \mathbf{H} = -\sum_{\ell=1}^{L-1}(\tch\cd{\ell+1}\cn{\ell}+\hc)
               -(\ttbc\cd{1}\cn{L}+\hc) +U\sum_{\ell=1}^{L}\hnocc{\ell\ups}\hnocc{\ell\dos}-\mu\sum_{\ell=1}^{L}\hnocc{\ell},
\end{align} where $\ttbc$ depends on the boundary condition. The Fermi operator
$\cd{\ell}$ creates an electron at site $\ell$. The symbols $\hnocc{\ell\mu}$
($\mu=\ups,\dos$) denote the number $\hnocc{\ell\mu}\equiv \cd{\ell\mu}\cn{\ell\mu}$ of
$\mu$-spin electrons at site $\ell$, and $\hnocc{\ell}\equiv
\hnocc{\ell\ups}+\hnocc{\ell\dos}$ denotes the site occupation. Sums over the
spin-component index $\sigma=\uparrow,\downarrow$ are implicit in the first, second, and
fourth terms on the right-hand side.

The fourth term introduces the chemical potential $\mu$, which controls the number of
electrons in the ground state. For fixed number $N$ of electrons, this term is a constant, which merely shifts the ground-state
energy and could have been left out. We nonetheless prefer to include it in the definition
of the Hamiltonian because attention to the chemical potential will prove instructive (see
Section~\ref{sec:25}, in particular).

\begin{figure}[h!]
  \centering
  \includegraphics[height=0.5\textheight]{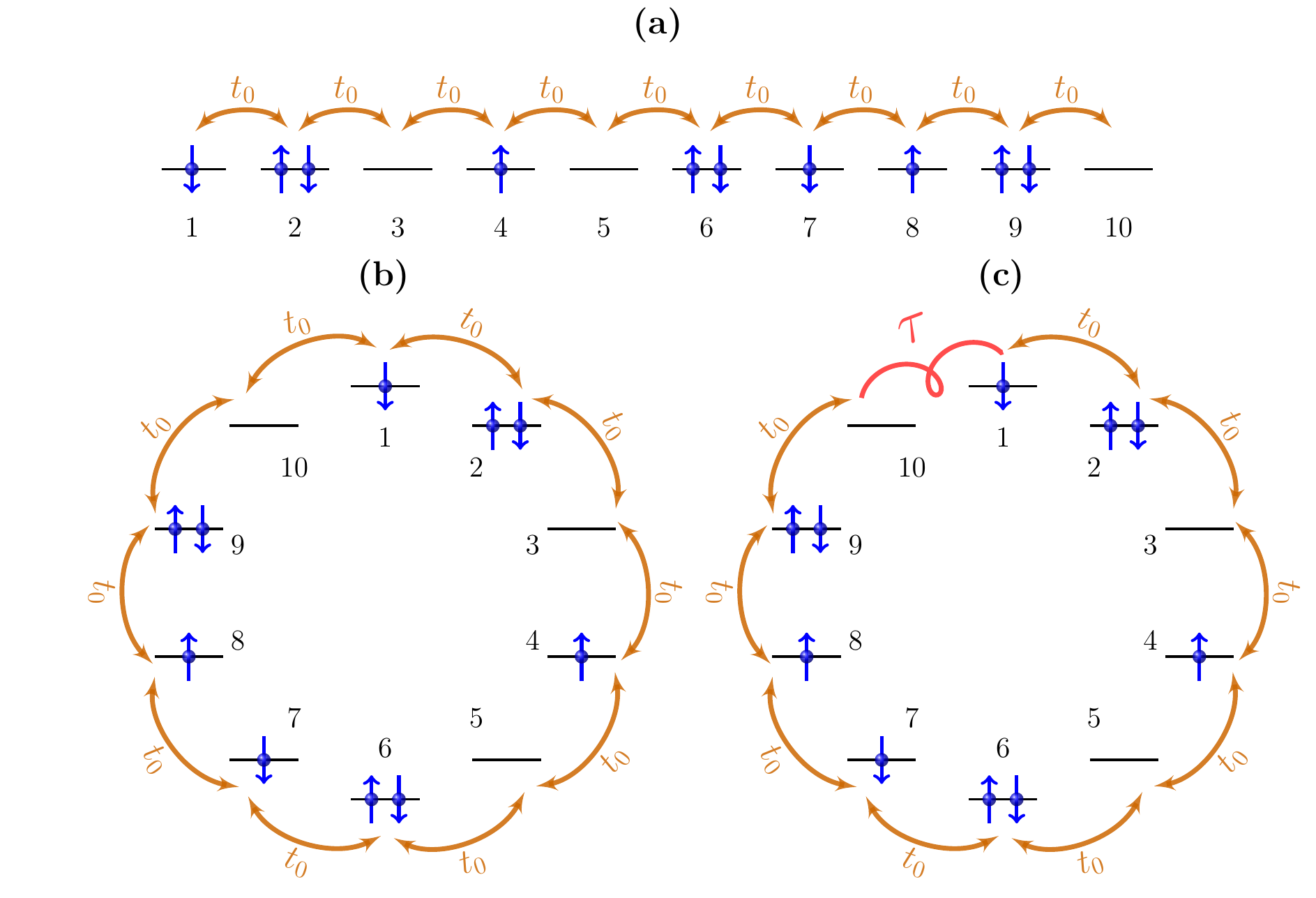}
  \caption{Boundary conditions. The three panels display the couplings in a ten-site
  Hubbard lattice under
    (a) open, (b) periodic, and (c) twisted boundary conditions.}
  \label{fig:1}
\end{figure}

As explained, we will discuss open, periodic, and twisted boundary conditions. The
coupling $\tau$ between the first and last chain sites specifies these conditions:
\begin{align}
  \label{eq:2} \ttbc =
  \begin{cases}
    0 & \mbox{open}\\ \tch & \mbox{periodic}\\ \tch e^{i\Theta}& \mbox{twisted}
  \end{cases},
\end{align} where the torsion $\Theta$ is an arbitrary real number. Of course, $\Theta$ is
only defined modulo $2\pi$. For $\Theta=0$, \tbc\ is equivalent to \pbc. $\Theta=\pi$
defines antiperiodic boundary condition, of secondary importance in our discussion.
Figure \ref{fig:1} schematically depicts the couplings under \obc, \pbc, and \tbc\ for
$L=10$.

As $L\to\infty$, the physical properties of the model become independent of boundary
condition. For small $L$ on the contrary, the properties are markedly affected by the
option on the right-hand side of Eq.~\eqref{eq:2}. Even the symmetry of the Hamiltonian is
affected, as detailed in the following section.

\subsection{Symmetry}
\label{sec:6} In the thermodynamical limit, i.~e., for $L\to\infty$, the Hubbard model
possesses a number of symmetries. Of special importance to our discussion are the
invarances under gauge transformation, rotation, particle-hole inversion, translation, and
mirror reflection. For finite $L$, the latter three depend on boundary condition. An
itemized discussion of the symmetries seems therefore appropriate.

\subsubsection{Global gauge transformation}
\label{sec:3} Inspection of Eq.~\eqref{eq:1} shows that the Hamiltonian remains invariant
under the global gauge transformation
\begin{align}
  \label{eq:3} \cn{\ell}\to e^{i\varphi}\cn{\ell},
\end{align} where $\varphi$ is a real constant.

Global gauge invariance is equivalent to charge conservation
\begin{align}
  \label{eq:4} [\mathbf{H},\mathbf{q}] =0,
\end{align} where $\mathbf{q}=\sum_{\ell}\hnocc{\ell}$.

That Eqs.~\eqref{eq:3}~and \eqref{eq:4} must be related follows from simple
considerations. For example, let us examine the first term on the \rhs\ of
Eq.~\eqref{eq:1} under \pbc. The product $\cd{1}\cn{L}$ will only remain invariant under
Eq.~\eqref{eq:3} if both operators, $\cd{1}$ and $\cn{L}$, undergo the same
transformation. If we apply the gauge-transformation~\eqref{eq:3} to the entire lattice
($\ell=1,\ldots, L$), the terms proportional to $\tau$ will be invariant. At the same
time, charge is conserved, because an electron can only hop from one site to another, both
within the lattice.

Let us now split the lattice in two sublattices, one comprising sites $\ell=1,2,\ldots, L-1$ and the
other, site $\ell=L$. If we apply the gauge-transformation~\eqref{eq:3} to the former, but not to
the latter, the terms proportional to $\ttbc$ on the \rhs\ of Eq.~\eqref{eq:1} will acquire phases.
The Hamiltonian will hence be modified. At the same time, charge will not be
conserved within each sublattice, since electrons can hop from one to the other.

As this simple example indicates, gauge invariance and charge conservation are intimately
related.  In fact, they are equivalent. The proof considers model Hamiltonians analogous
to Eq.~\eqref{eq:1}, comprising terms such as the ones on the \rhs, of the general form
\begin{align}
  \label{eq:5} \hat h= \sum_{\substack{m\sub1,\ldots m\sub M,\\p\sub1,\ldots p\sub
  P=1}}^{L}
A_{m\sub1\ldots m\sub M}^{p\sub1\ldots p\sub P}
 \cd{m\sub1}\ldots\cd{m\sub M}\cn{p\sub1}\ldots\cn{p\sub{P}},
\end{align} where $M$ and $P$ are integers. For instance, $M=P=2$ in the Coulomb-repulsion
term on the \rhs\ of Eq.~\eqref{eq:1}, while $M=P=1$ in the other terms.

Under Eq.~\eqref{eq:3}, the Hamiltonian~\eqref{eq:5} transforms as
\begin{align}
  \label{eq:6} \hat h\to e\super{i(P-M)\varphi}\sum_{\substack{m\sub1,\ldots m\sub
  M,\\p\sub1,\ldots p\sub P=1}}^{L} A_{m\sub1\ldots m\sub M}^{p\sub1\ldots p\sub P}
  \cd{m\sub1}\ldots\cd{m\sub M}\cn{p\sub1}\ldots\cn{p\sub{P}},
\end{align} and hence remains invariant if and only if $M=P$.

Likewise, charge is conserved if and only if $M=P$. To prove that, it is expedient to
evaluate the commutator
\begin{align}
  \label{eq:7} [\hat h,q] = \sum_{\ell}\Big([\hat h,\cd{\ell}]\cn{\ell}+\cd{\ell}[\hat
  h,\cn{\ell}]\Big).
\end{align}

Computation of each commutator on the \rhs\ of Eq.~\eqref{eq:7} shows that
\begin{align}
  \label{eq:8} [\hat h,\cd{\ell}]\cn{\ell} = \sum_{\substack{m\sub1,\ldots m\sub
  M,\\p\sub1,\ldots p\sub P=1}}^{L} \Big[A_{m\sub1\ldots m\sub M}^{p\sub1\ldots p\sub P}
 \cd{m\sub1}\ldots\cd{m\sub M}\cn{p\sub1}\ldots\cn{p\sub{P}}
                                           (\delta_{\ell,p\sub1}+\ldots+\delta_{\ell,p\sub P})\Big],
\end{align} and
\begin{align}
  \label{eq:9}
    \cd{\ell}[\hat h,\cn{\ell}] = -\sum_{
  \substack{m\sub1,\ldots m\sub M,\\p\sub1,\ldots p\sub P=1}}^{L}
  \Big[A_{m\sub1\ldots m\sub M}^{p\sub1\ldots p\sub P}
         \cd{m\sub1}\ldots\cd{m\sub M}\cn{p\sub1}\ldots\cn{p\sub{P}}
  (\delta_{\ell,m\sub1}+\ldots+\delta_{\ell,m\sub M})\Big],
\end{align} and therefore
\begin{align}
  \label{eq:10} [\hat h,q] = (P-M)\hat h,
\end{align} which shows that $[\hat h,q]=0$ if and only if $P=M$.

Since each term on the \rhs\ of Eq.~\eqref{eq:1} is unaffected by the
transformation~\eqref{eq:3}, the Hubbard Hamiltonian is gauge invariant and conserves
charge under any of the boundary conditions in Eq.~\eqref{eq:2} To reach the same
conclusion in an alternative way, we only have to compute the commutator on the left-hand
side of Eq.~\eqref{eq:4}, which yields zero.

\subsubsection{Local gauge transformation}
\label{sec:10} Unlike the global transformation in Eq.~\eqref{eq:3}, local gauge
transformations tend to modify the form of the Hamiltonian~\eqref{eq:1} Of special
interest is the transformation
\begin{align}
  \label{eq:11}
    \cn{\ell}\equiv e^{i\ell\alpha }\an{\ell}\qquad(\ell=1,\ldots,L),
\end{align} where $\alpha$ ($0\le\alpha<2\pi$) is a constant, so that the phase
$\ell\alpha$ grows uniformly along the lattice.

Substitution of the \rhs\ of Eq.~\eqref{eq:11} for the $\cn{\ell}$ in Eq.~\eqref{eq:1}
yields the expression
\begin{align}
  \label{eq:12}
    \mathbf{H} &= -\sum_{\ell=1}^{L-1}(\tch e^{-i\alpha}\ad{\ell+1}\an{\ell}+\hc)
           -(\ttbc e^{i(L-1)\alpha}\ad{1}\an{L}+\hc)
                 +U\sum_{\ell=1}^{L}\hbnocc{\ell\ups}\hbnocc{\ell\dos}-\mu\sum_{\ell=1}^{L}\hbnocc{\ell},
\end{align} where $\hbnocc{\ell}\equiv \ad{\ell}\an{\ell}$.

If $\tch$ is a complex number with phase $\beta$, i.~e., if $\tch=|\tch|e^{i\beta}$, we
can choose $\alpha=\beta$ to make real the coefficients $\tch e^{-i\alpha}$ and $\tchs
e^{i\alpha}$ on the \rhs\ of Eq.~\eqref{eq:12}. The torsion $\Theta$ is then transformed to
\begin{align}
  \label{eq:13} \Theta' =\Theta+L\beta.
\end{align}

With no loss of generality, therefore, we can take the coefficients $\tch$ on the \rhs\ of
Eq.~\eqref{eq:1} to be real and will do so henceforth.

\subsubsection{Rotation in spin space}
\label{sec:4} Clearly, the Hamiltonian~\eqref{eq:1} remains invariant under the
spin-component transformation $\cn{\ell\,\sigma}\to\cn{\ell\,-\sigma}$
($\sigma=\uparrow,\downarrow$). More generally, it possesses SU(2) symmetry in spin space
and hence conserves spin. The boundary term
$(\ttbc\cd{1\uparrow}\cn{L\uparrow}+\ttbc\cd{1\downarrow}\cn{L\downarrow}+\hc)$ is
likewise symmetric and conserves spin, for \obc, \pbc, or \tbc.

\subsubsection{Inversion}
\label{sec:9} The last two terms on the right-hand side of Eq.~\eqref{eq:1} remain
invariant under the transformation $\ell\to L+1-\ell$ ($\ell=1,2,\ldots,L$), which
reverses the ordering of the lattice sites.  Whether the first and second terms also
remain invariant is less evident. Define, therefore, the Fermi operators
\begin{align}
  \label{eq:14} \an{L+1-\ell} \equiv \cn{\ell}\qquad(\ell=1,2,\dots,L).
\end{align}

Substitution of the $\an{L+1-\ell}$ for the $\cn{\ell}$ on the \rhs\ of Eq.~\eqref{eq:1}
expresses the model Hamiltonian on the basis of the former:
\begin{align}
  \label{eq:15} \mathbf{H} &= -\sum_{\ell=1}^{L-1}\tch(\ad{L-\ell}\an{L+1-\ell}+\hc)
           -(\ttbc\ad{L}\an{1}+\hc)
 +U\sum_{\ell=1}^{L}\hbnocc{L+1-\ell\uparrow}\hbnocc{L+1-\ell\downarrow}
-\mu\sum_{\ell=1}^{L}\hbnocc{L+1-\ell}
\end{align}

We then relabel the summation indices on the \rhs\ of Eq.~\eqref{eq:15}, letting $\ell\to
L-\ell$ in the first sum, and $\ell\to L+1-\ell$ in the third and fourth ones, to show
that
\begin{align}
  \label{eq:16} \mathbf{H} &=
  -\sum_{\ell=1}^{L-1}\tch(\ad{\ell}\an{\ell+1}+\ad{\ell+1}\an{\ell})
           -(\ttbc\ad{L}\an{1}+\ttbcs\ad{1}\an{L})
+U\sum_{\ell=1}^{L}\hbnocc{\ell\uparrow}\hbnocc{\ell\downarrow}-\mu\sum_{\ell=1}^{L}\hbnocc{\ell},
\end{align} where we have spelled out the second terms within the parentheses on the \rhs\
to recall that $\tch$ is real, while $\ttbc$ may be complex.

The first, third, and fourth terms on the \rhs\ of Eq.~\eqref{eq:16} are equivalent to the
corresponding terms on the \rhs\ of Eq.~\eqref{eq:1}. The second term, however, is
equivalent to the Hermitian conjugate of the second term on the \rhs\ of Eq.~\eqref{eq:1}.
In other words, inversion maps $\ttbc$ onto $\ttbcs$. As long as $\ttbc$ is real, i.~e.,
for \obc\ ($\ttbc=0$), \pbc\ ($\ttbc=\tch$) or for anti-periodic boundary condition
($\ttbc=-\tch$), we can see that $\mathbf{H}$ remains invariant under inversion. Twisted
boundary condition breaks inversion symmetry, except for $\Theta=0\mod\pi$.

\subsubsection{Particle-hole transformation}
\label{sec:8} The standard electron-hole transformation, which exchanges the roles of
filled states below the Fermi level and vacant states above the Fermi level, merely shifts
the chemical potential of the infinite-lattice Hubbard Hamiltonian, from $\mu$ to
$U-\mu$.\cite{2003EFG+} If $\mu=U/2$, the Hamiltonian remains invariant. 
Extensions to
finite lattices calls for special attention to boundary condition, as shown next.

We start with the equality defining the conventional electron-hole transformation:
\begin{align}
  \label{eq:17} \an{\ell} \equiv (-1)^{\ell}\cd{\ell}.
\end{align}

Substitution of Eq.~\eqref{eq:17} for the Fermi operators on the \rhs\ of Eq.~\eqref{eq:1}
shows that
\begin{align}
  \label{eq:18} \mathbf{H} &= \sum_{\ell=1}^{L-1}\tch(\an{\ell+1}\ad{\ell}+\hc)
           +(-1)^{L}(\ttbc\an{1}\ad{L}+\hc)
  +U\sum_{\ell=1}^{L}(1-\hbnocc{\ell\ups})(1-\hbnocc{\ell\dos})-\mu\sum_{\ell=1}^{L}(2-\hbnocc{\ell}),
\end{align} where $\hbnocc{\ell}\equiv \ad{\ell}\an{\ell}$.

We now bring the first two terms on the \rhs\ of Eq.~\eqref{eq:18} to normal order and
simplify the last two to obtain the expression
\begin{align}
  \label{eq:19} \mathbf{H} =& -\sum_{\ell=1}^{L-1}\tch(\ad{\ell}\an{\ell+1}+\hc)
                -(-1)^{L}(\ttbcs\ad{1}\an{L}+\ttbc\ad{L}\an{1})\nc
              &+(U-2\mu)L+U\sum_{\ell=1}^{L}\hbnocc{\ell\ups}\hbnocc{\ell\dos}
                -(U-\mu)\sum_{\ell=1}^{L}\hbnocc{\ell}.
\end{align}

The third term on the \rhs\ of \eqref{eq:19} is a constant that merely shifts the zero of
energy. We leave it aside and compare the other terms with those on the \rhs\ of
Eq.~\eqref{eq:1}. The first terms on the {\rhs}s of the two equalities and the terms
proportional to $U$ have the same form. Comparison between the last terms shows that the
particle-hole inversion maps $\mu\to U-\mu$. These conclusions are independent of boundary
condition and lattice size.
By contrast, the second term on the {\rhs} of Eq.~\eqref{eq:19}, which enforces boundary
condition, is a function of $L$. Equivalence with the corresponding term on the \rhs\ of
Eq.~\eqref{eq:1} is insured if and only if
\begin{align}
  \label{eq:20} \ttbc = \ttbcs(-1)^{L}.
\end{align}

Under \obc, $\ttbc=\ttbcs=0$, and Eq.~\eqref{eq:20} is always satisfied. Under \pbc,
$\ttbc=\ttbcs=\tch$, and it follows that Eq.~\eqref{eq:20} is only satisfied for even
$L$. Finally, under \tbc, $\ttbc=\tch\exp(i\Theta)$, while $\ttbcs=\tch\exp(-i\Theta)$,
and it follows that Eq.~\eqref{eq:20} is equivalent to the condition
\begin{align}
  \label{eq:21} \Theta =\dfrac{\pi}2L\mod{\pi}.
\end{align}

Given that $\Theta$ is only defined modulo $2\pi$, we can see that Eq.~\eqref{eq:20} is
equivalent to the requirement that $\Theta$ be either 0 or $\pi$ for even $L$ and
$\Theta=\pm\pi/2$ for odd $L$.  For the illustrative purposes of our discussion, it is
more convenient to consider the sufficient condition
\begin{align}
  \label{eq:22} \Theta = \dfrac{\pi}2L,
\end{align} which can be spelled out as follows:
\begin{align}
  \label{eq:23} \Theta=
  \begin{cases}
    0&\qquad(L=4\ell)\\ \dfrac{\pi}2&\qquad(L=4\ell+1)\\ \pi&\qquad(L=4\ell+2)\\
    -\dfrac{\pi}2&\qquad(L=4\ell+3)
  \end{cases},
\end{align} where $\ell=0,1,2\ldots$. With $\Theta=0$ ($\Theta=\pi$) the model is under
periodic (anti-periodic) boundary condition.

As long as Eq.~\eqref{eq:22} is satisfied, Eq.~\eqref{eq:19} reads
\begin{align}
  \label{eq:24} \mathbf{H} =& -\sum_{\ell=1}^{L-1}\tch(\ad{\ell+1}\an{\ell}+\hc)
                -(\ttbc\ad{1}\an{L}+\hc)+(U-2\mu)L+U\sum_{\ell=1}^{L}\hbnocc{\ell\ups}\hbnocc{\ell\dos}
    -(U-\mu)\sum_{\ell=1}^{L}\hbnocc{\ell}.
\end{align}

With the substitution $\mu\to U-\mu$, Eq.~\eqref{eq:24} reproduces Eq.~\eqref{eq:1}. For
$\mu=U/2$, in particular, the {\rhs} remains invariant under particle-hole
transformation. Equation~\eqref{eq:22} therefore insures particle-hole symmetry.

\subsubsection{Translation}
\label{sec:7} The last two terms on the right-hand side of Eq.~\eqref{eq:1} are invariant
under the transformation
\begin{subequations}
  \label{eq:25}
  \begin{align}
    \label{eq:26} \ell\to \ell+1&\qquad(\ell=1,2,\ldots,L-1)\\ \ell\to 1
    &\qquad(\ell=L).\label{eq:27}
  \end{align}
\end{subequations} The first term on the right-hand of Eq.~\eqref{eq:1}, however, is
modified by the same transformation. With $\ttbc=\tch$ (\pbc), the sum of the first and
second terms remains invariant. For any $L$, therefore, under \pbc, the Hamiltonian is
translationally invariant. With $\ttbc=0$ (\obc), by contrast, translational symmetry is
lost. At first sight, \tbc\ may seem to also break translational invariance, but the
following reasoning leads to the opposite conclusion.

Given a torsion $\Theta$, define the \emph{local torsion}
\begin{align}
  \label{eq:28} \theta \equiv \dfrac{\Theta}L
\end{align} and the Fermi operators
\begin{align}
  \label{eq:29} \an{\ell}\equiv \cn{\ell}e^{i\ell\theta},
\end{align} so that
$\ad{\ell\sigma}\an{\ell\sigma}=\cd{\ell\sigma}\cn{\ell\sigma}\equiv\hnocc{\ell\sigma}$.

Equation~\eqref{eq:1} can then be written in the form
\begin{align}
  \label{eq:30} \mathbf{H} &= -\sum_{\ell=1}^{L-1}(\tch e^{i
  \theta}\ad{\ell+1}\an{\ell}+\hc)
               +U\sum_{\ell=1}^{L}\hnocc{\ell\uparrow}\hbnocc{\ell\downarrow}-(\tch
           e^{iL\theta}e^{i(1-L)\theta}\ad{1}\an{L}+\hc)-\mu\sum_{\ell=1}^{L}\hbnocc{\ell},
\end{align} which simplifies to the expression
\begin{align}
  \label{eq:31} \mathbf{H} = -\sum_{\ell=1}^{L}(\tun\ad{\ell+1}\an{\ell}+\hc)
+U\sum_{\ell=1}^{L}\hbnocc{\ell\uparrow}\hbnocc{\ell\downarrow}
-\mu\sum_{\ell=1}^{L}\hbnocc{\ell},
\end{align} where we have defined the complex coupling $\tun\equiv\tch e^{i\theta}$ and
identified $\an{L+1}$ with $\an{1}$.

Equation~\eqref{eq:31} is equivalent to Eq.~\eqref{eq:1} with $\ttbc=\tun$. Moreover, its
right-hand side remains invariant under the lattice
translations~\eqref{eq:26}~and~\eqref{eq:27}. The one-dimensional Hubbard Hamiltonian
under \pbc\ or \tbc\ is therefore covered by Bloch's Theorem.\cite{Ashcroft_1976} The
discussion in Sec.~\ref{sec:18} will benefit from the ensuing momentum-conservation law.

Table~\ref{tab:1} summarizes the properties of the model under inversion, translation, and
particle-hole transformation.
\begin{table}[th!]
  \centering
  \begin{tabular}{lrrccc}
    \hline\hline \textbf{BC}& \textbf{Transform}& \multicolumn{1}{c}{$L$} &
    \textbf{Invariant?}\footnote{At
                                                               half-filling, i.~e.,
    $\mu=U/2$}&$\Theta'$&$\mu'$\\ \hline \textbf{\obc} & inversion & any & yes & $-$
    &$\mu$ \\
               & translation & any & no & $-$ & $-$ \\ & p-h & any & yes & $-$ & $U-\mu$
               \\[1mm]
    \textbf{\pbc} & inversion &any & yes & $\Theta$ & $\mu$\\
               & translation & any & yes & $\Theta$ & $\mu$\\ & p-h & even & yes & 0 &
               $U-\mu$ \\ & p-h & odd & no & $\pi$ & $U-\mu$ \\[1mm]
    \textbf{\tbc}\footnote{Except $\Theta=(\pi/2)L$} & inversion & any & no & $-\Theta$ &
    $\mu$ \\
               & translation& any & yes & $\Theta$ & $\mu$\\ & p-h & any & no & $\pi
               L-\Theta$ & $U-\mu$ \\[1mm]
    $\mathbf{\Theta}=\dfrac{\pi}{2}L$ & inversion & even & no & $\Theta\mod 2\pi$ & $\mu$
    \\[2mm] & inversion & odd & no & $-\Theta$ & $\mu$ \\[2mm]
               & translation & any & yes & $\Theta$ & $\mu$\\[2mm] & p-h & any & yes &
               $\Theta$ & $U-\mu$ \\[2mm]
    \hline\hline
  \end{tabular}
  \caption[behave]{Behavior of the finite-size Hubbard Hamiltonian under left-right
  inversion,
    translation, and particle-hole transformation. Open, periodic, or twisted boundary
    conditions are considered, with even or odd number $L$ of sites. The symbol '$-$'
    indicates that the corresponding parameter is undefined.  Under particle-hole
    transformations, the Coulomb chemical potential $\mu$ is mapped onto $U-\mu$, while
    the ground-state energy is shifted by $\mu'-\mu\equiv U-2\mu$.  For convenience, the
    last four rows describe the model under twisted boundary condition with the special
    torsion $\Theta=(\pi/2)L$, which is particle-hole symmetric at half filling for any
    $L$.}
  \label{tab:1}
\end{table}

\section{Analytical results}
\label{sec:18} We are interested in the physical properties of the finite-size
unidimensional Hubbard model under different boundary conditions. Numerical results for the
ground-state energy, electronic density and magnetization, and for the energy gap of the small-$L$
Hamiltonian at half filling will be discussed in Section~\ref{sec:17}.  Preparatory to that
discussion and to gain preliminary physical insight, we survey analytical expressions
covering special limits, leaving more detailed discussion to the appendices.  For the
uncorrelated model ($U=0$), Appendix~\ref{sec:29} identifies the dispersion relation pertaining to
each boundary condition, from which the ground-state energy and gap can be easily obtained, and also
discusses the electronic and magnetization densities. For $U>0$, Appendix~\ref{sec:30} recapitulates
results extracted from the Bethe-Ansatz diagonalization of the model Hamiltonian, which become
simple only in the $U\to\infty$ limit.

\subsection{$U=0$}
\label{sec:19}

With $U=0$ the Hamiltonian~\eqref{eq:1} becomes quadratic. We can easily diagonalize it, under \pbc,
\tbc, or \obc. Since Bloch's Theorem covers only the former two boundary conditions, however,
Appendix~\ref{sec:29} follows distinct procedures and obtains distinct results, depending on whether
one is dealing with closed (\pbc\ or \tbc) or \obc. The results are summarized in
Table~\ref{tab:2}. In the infinite model, the per-particle ground-state energy is
\begin{align}
  \label{eq:124}
    E\sub{\Omega}=-\dfrac{4\tch}{\pi}\approx-1.27\tch.
\end{align}
The same result can be obtained from the $L\to\infty$ limit of each expression for the ground-state
energy in the table.

\begin{table}[h!]
  \centering
  \begin{tabular}{rrcl}
    \hline\hline \textbf{Boundary} \\ \textbf{Condition} & $L$ &\,&
    $-E\sub{\Omega}/(2\tch)$\\[2mm] \hline\\[-2mm] \textbf{Open} & even &&
    $\dfrac1{\sin(\frac{\pi}{2(L+1)})}-1$\\[4mm] & odd &&
    $\dfrac1{\tan(\frac{\pi}{2(L+1)})}-1$\\[4mm] \textbf{Periodic} & $4n$ && $
    \dfrac2{\tan(\frac{\pi}{L})}$\\[4mm]
     & $4n+2$ && $ \dfrac2{\sin(\frac{\pi}{L})}$\\[4mm] & odd && $
     \dfrac{\cos(\frac{\pi}{2L})}{\tan(\frac{\pi}{2L})}$\\[4mm]
    \textbf{Twisted} & even && $ \dfrac2{\tan(\frac{\pi}L)}$\\[4mm] & odd &&
    $\dfrac1{\tan(\frac{\pi}{2L})}$\\[4mm] \hline\hline
  \end{tabular}
  \caption{Ground-state energies for $U=0$ and $N=L$ under open, periodic, and twisted
  [$\Theta=(\pi/2)L$]
    boundary conditions. }
  \label{tab:2}
\end{table}

\begin{figure}[h!]
  \centering
  \includegraphics[width=\columnwidth]{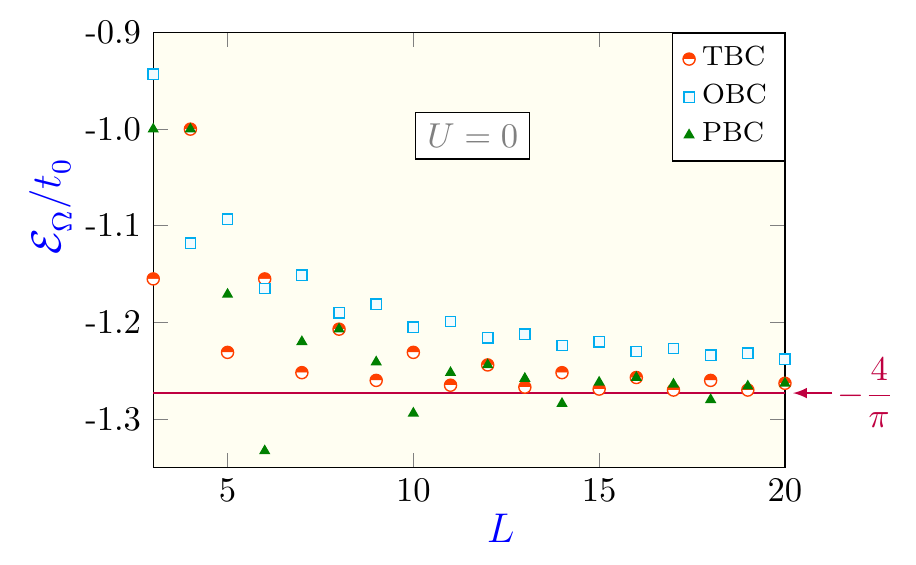}
  \caption[\ ]{Per-particle ground-state energies for $U=0$ under \obc\ (open squares), \pbc\
    (filled triangles), and \tbc\ with the special torsion $\Theta=(\pi/2)L$ (half-filled
    circles). To avoid compression of the vertical axis, we have left out the $L=2$ data,
    for which the per-particle ground-state energy vanishes under \tbc. The horizontal,
    magenta solid line shows the $L\to\infty$ limit, Eq.~\eqref{eq:124}. For $L=4,8,12,16,$
    and 20, i.~e., for multiples of four, the filled triangles and half-filled circles
    coincide. Under \tbc, the per-site energies for $L=3, 5,7$, and 9 are equal to the
    per-site energies for $L=6,10,14$, and 18, respectively.}
  \label{fig:3}
\end{figure}

Results for $3\le L \le30$ are displayed in Fig.~\ref{fig:3}. The arrow pointing to the right-hand
vertical axis shows that, as the lattice size $L$ grows, the three sets of data representing \tbc\
with $\Theta=(\pi/2)L$ (half-filled circles), \pbc\ (filled triangles), and \obc\ (open squares)
approach $-(4/\pi)\tch$, the per-particle ground-state energy for $L\to\infty$.
The convergence is staggered, rather than smooth, and boundary-condition dependent. The
open squares representing \obc\ stagger the least, but converge relatively slowly to the
horizontal line marking the infinite-lattice limit. Periodic boundary condition ensures
faster convergence, but the filled triangles for $L=4n+2$ ($n=1,2,3,$ and 4) lie below the
horizontal line, while the triangles for the other lattice sizes lie above it.
Finally, under \tbc, the per-particle energies $\mathcal{E}\sub{\Omega}\equiv
E\sub{\Omega}/N$ decay rapidly to the horizontal line. For $L=4\ell$ ($\ell=1,2,3,4,$ and
5), the half-filled circles coincide with the filled triangles, as one would expect from
Table~\ref{tab:2} or from recalling from Eq.~\eqref{eq:23} that $\Theta=(\pi/2)L$ is
equivalent to $\Theta=0$ when $L$ is a multiple of four.

The sequence of odd-$L$ half-filled circles show especially rapid convergence. In fact,
comparison of the last two rows in Table~\ref{tab:2} shows that the per-particle energies
at half-filling for odd lattice size $L$ coincide with the per-particle energies at
half-filling for lattice size $2L$. The even $L$ convergence is substantially slower,
since, as the figure shows, the deviation from the $L\to\infty$ limit for $L=3, 5, 7$, and
$9$, for instance, are equal to the deviations for $L=6, 10, 14$, and 18,
respectively. Section~\ref{sec:24} will discuss this coincidence further.

\subsection{Density and magnetization density}
\label{sec:31}

Other ground-state properties of interest are the electronic density
$n\sub{\ell}=n\sub{\ell\ups}+n\sub{\ell\dos}$ ($\ell=1,\ldots,L)$ and magnetization
density $m\sub{\ell}=n\sub{\ell\ups}-n\sub{\ell\dos}$, two functions of paramount
importance in Density Functional Theory.\cite{[{See, for instance,\ }]1983Kohn79}

As explained by Sec.~\ref{sec:22}, the electronic density at half-filling is uniformly unitary for
all $U$ and $L$. When $L$ is even, the magnetization density vanishes for all $U$. For finite, odd
$L$, however, the magnetization density is nonzero and must be computed numerically for $U\ne0$. An
exception is the $U\to\infty$ limit of the $L=3$ model, which yields analytical results.

In the large $U$ limit, the charge degrees of freedom being frozen at $n\sub{\ell}=1$
($\ell=1,2,3$), each site is equivalent to a spin-$1/2$ variable\textemdash a
doublet. There are, therefore, $2^{3}=8$ states, which can be classified by the total spin
$S$, because as explained in Sec.~\ref{sec:4}, $S$ is conserved.

The total spin resulting from the addition of three individual spins can either be $S=3/2$
or $S=1/2$. The quadruplet ($S=3/2$) comprises four of the eight states; the other four
must belong to two doublets ($S=1/2$).

Consider the $S\sub{z}=1/2$ components of the two $S=1/2$ states. Since they have the same
spin, we are free to choose any pair of orthonormal states that are orthogonal to the
$S\sub{z}=1/2$ component of the triplet. The latter has the expression
\begin{align}
  \label{eq:105} \ket{S=\dfrac32,S\sub{z}=\dfrac12}=\dfrac{1}{\sqrt3}
\Big(\cdu{1}\cdu{2}\cdd{3}+\cdu{1}\cdd{2}\cdu{3}+\cdd{1}\cdu{2}\cdu{3}\Big)\kvac.
\end{align}

Two convenient choices for $S\sub{z}=S=1/2$ are
\begin{align}
  \label{eq:106} \ket{\dfrac12,\dfrac12,u} =
  \dfrac{1}{\sqrt2}\Big(\cdu{1}\cdu{2}\cdd{3}-\cdd{1}\cdu{2}\cdu{3}\Big)\kvac,
\end{align} which is odd ($u$) under spatial inversion, and
\begin{align}
  \label{eq:107} \ket{\dfrac12,\dfrac12,g} =
\dfrac{1}{\sqrt6}\Big(\cdu{1}\cdu{2}\cdd{3}-2\cdu{1}\cdd{2}\cdu{3}+\cdd{1}\cdu{2}\cdu{3}\Big)\kvac,
\end{align} which is even ($g$).

Straightforward computation shows the {\rhs}s of Eqs.~\eqref{eq:106}~and \eqref{eq:107} to
be orthogonal to the \rhs\ of Eq.~\eqref{eq:105}. In addition, since they have opposite
parities, $\ket{1/2,1/2,u}$ and $\ket{1/2,1/2,g}$ are mutually orthogonal.

For infinite $U$, the quadruplet and the two doublets are degenerate, with zero
energy. For large, finite $U$, however, the kinetic terms in the model Hamiltonian can
contribute energies of the order of $-t_{0}^{2}/U$. From Eq.~\eqref{eq:105} we find that
\begin{align}
  \label{eq:108} \mathbf{H}\ket{\dfrac32,\dfrac12}=0,
\end{align} which shows that $\ket{\frac32,\frac12}$ is an eigenstate with zero energy,
for all $U$. In fact, given spin-rotation symmetry, it shows that each component of the
quadruplet is an eigenstate, with $E=0$. Second-order perturbation theory \cite{[{See, for
instance,\ }]1994Griffiths} on the other hand shows that, for large $U/\tch$, the two
doublet components in Eqs.~\eqref{eq:106}~and \eqref{eq:107} have negative energies that
differ by $\mathcal{O}(t_{0}^2/U)$, the even combination $\ket{\frac12,\frac12,g}$ being
the ground state.

From Eq.~\eqref{eq:107}, we can now compute the magnetization density for
$\ket{1/2,1/2,g}$:
\begin{align}
  \label{eq:109}
    m^{g}_{\ell}=
  \begin{cases}
   \ \ \dfrac23&\qquad(\ell=1,3)\\[3mm]
    -\dfrac13&\qquad(\ell=2)
  \end{cases}.
\end{align}

Neither this attractively simple result, nor the simple analysis leading to it can be
extended to $N=L>3$. As the lattice becomes larger, the number of spin states grows
exponentially, and so does the dimension of the $Q,S, S\sub{z}, \Pi$ (where $\Pi$ denotes
parity under lattice inversion) sector containing the ground state. Already for $L=7$ the
matrix resulting from the projection of the Hamiltonian is too big to be analytically
diagonalized, and numerical treatment becomes necessary.

\subsection{$U\to\infty$}
\label{sec:20}
The Coulomb repulsion $U$ penalizes double occupation of the $\cn{\ell}$ orbitals. The
eigenstates of the $U\ne0$ model Hamiltonian are no longer mutually independent, and the
single-particle description breaks down. As $U\to\infty$, the energetic cost of double
occupation becomes prohibitive and, for $N\le L$, each orbital $c\sub{\ell}$
($\ell=1,\ldots,L$) can hold no more than one electron.  In this limit, in analogy with
the depictions in Fig.~\ref{fig:2}, one might hope to recover a simple picture of the
ground state comprising $L$ levels labeled by momenta $k$. The lowest $N$ levels would
then be singly occupied, and the remaining $L-N$ ones would be empty.

This description is ratified by the Bethe-Ansatz solution,\cite{1968LiW1445,2003LiW1} but
the computation of the allowed momenta requires special attention. Under \obc\
Eq.~\eqref{eq:50} is still valid. Under \pbc\ or \tbc, however, the conditions determining
the allowed $k$ depend not only on $L$, but also on the ground-state spin $S$ and its
component $S\sub{z}$. Given this distinction, Appendix~\ref{sec:30} discusses open and closed
(\pbc\ or \tbc) boundary conditions under separate headings.

Under \obc, the computation of ground-state energies is relatively simple (see
Appendix~\ref{sec:14}). For closed boundary conditions, however, one must refer to the Bethe-Ansatz
solution. The procedure developed by Lieb and Wu\cite{1968LiW1445,2003LiW1} yields two sets of exact
nonlinear equations \textemdash the Lieb-Wu Equations\textemdash that determine the ground-state
energy. In most cases, these equations yield only to numerical treatment. In the $U\to\infty$ limit,
however, the two sets of Lieb-Wu Equations can be uncoupled, one of them being mapped onto a gas of
noninteracting particles, as detailed in Appendix~\ref{sec:35}. 

As illustrations, Table~\ref{tab:3} shows the resulting ground-state energies (shifted by $\mu N$)
for $N=L-1$ for $L=2,\ldots,10$. In all rows, the energy is $E_{\Omega}= -2\tch\sin(k)$,
where $k$ is either $\pi/2$ or a multiple of $2\pi/NL$ that is close to $\pi/2$. The
ground state is degenerate. In particular, its spin can have multiple values.  The
ground-state spin is $S=N/2$ if and only if $L$ is a multiple of four. This result
contrasts with Nagaoka's theorem,\cite{1966Nagaoka392,1965Nagaoka409,1989Tasaki9192} which
states that, for various two- or three-dimensional lattices, the $U\to\infty$ ground state
of the Hubbard Hamiltonian acquires the maximal spin $S=N/2$ at $N=L-1$, where $L$ is the
number of lattice sites.

\begin{table}[th!]
  \centering
  \begin{tabular}{cccc}
    \hline\hline $L$ & $N$ & $2S+1$ & $-(E\sub{\Omega}+\mu N)/2\tch$ \\[1mm] \hline 2 & 1
    & 2 & 0 \\ 3 & 2 &1,3& $\sin(\frac{\pi}3)$\\ 4 & 3 & 4 & 1\\ 5 & 4 & 3 &1\\ 6 & 5 &2,4
    & $\sin(\frac{8\pi}{15})$\\ 7 & 6 & 1,3,5 & $\sin(\frac{11\pi}{21})$\\ 8 & 7 &
    2,4,8&1\\ 9 & 8 & 1,3,5,7 &1\\ 10 & 9 & 2,4,6,8 &$\sin(\frac{23\pi}{45})$\\[1mm]
    \hline\hline
  \end{tabular}
  \caption{Ground-state energies for $U\to\infty$ Hubbard Hamiltonians with different
  lengths
    $L$ and twisted boundary conditions with torsion $\Theta=(\pi/2)L$, for $N=L-1$. The
    third column displays the ground-state spin multiplicities $2S+1$.}
  \label{tab:3}
\end{table}

\subsubsection{Density and magnetization density}
\label{sec:22}  While
the density and magnetization for the half-filled Hubbard chain under \pbc\ or \tbc, and
the density under \obc\ can be easily understood on the basis of symmetry, the
magnetization under \obc\ requires special discussion.

\paragraph{Periodic and twisted boundary conditions}
\label{sec:27} Under \pbc\ or \tbc, arbitrary lattice translations leave physical
properties unchanged. Both $n\sub{\ell}$ and $m\sub{\ell}$ must therefore be independent
of $\ell$, i.~e., uniform. At half-filling, with $N=L$, the density must be unitary,
$n\sub{\ell}=1$ ($\ell=1,2,\ldots,L$).

The magnetization density depends on the parity of $L$. For even $L$, the $N=L$ electrons
can be divided into $N/2$ $\ups$-spin and $N/2$ $\dos$-spin electrons. The ground state is
a singlet and the magnetization vanishes. It follows that $m\sub{\ell}=0$
($\ell=1,2,\ldots,L$). For odd $L$, the ground state is a doublet ($S=1/2$). If $S\sub
z=1/2$, the numbers of $\ups$-spin and $\dos$-spin electrons must be $N\sub{\ups}=(L+1)/2$
and $N\sub{\dos}=(L-1)/2$, respectively, and the resulting magnetization is $M=1$. The
magnetization density is therefore $m\sub{\ell}=1/L$ ($\ell=1,2,\ldots,L$).

\paragraph{Open boundary condition}
\label{sec:28} Under \obc\ translation invariance is broken, and one would expect the
density and the magnetization density to be position dependent. For $N=L$, particle-hole
symmetry nonetheless forces the density to be uniform, as a simple argument shows. Under
particle-hole transformation, the density $n\sub{\ell}$ at site $\ell$ is transformed to
$2-n\sub{\ell}$. The $N=L$ Hamiltonian being particle-hole symmetric, we can conclude that
$n\sub{\ell}=2-n\sub{\ell}$, and hence that $n\sub{\ell}=1$.

The magnetization density, on the other hand, may or may not be uniform, depending on the
parity of $N=L$. For even $L$, the numbers $N\sub{\ups}$ and $N\sub{\dos}$ of $\ups$- and
$\dos$-spin electrons in the ground state are equal, $N\sub{\ups}=N\sub{\dos}=N/2$. The
ground state is a singlet, hence invariant under the transformation
$S\sub{z}\to-S\sub{z}$, which turns the $\sigma$-spin density $n\sub{\sigma}$ into
$n\sub{-\sigma}$ ($\sigma=\ups,\dos$). It follows that $n\sub{\ups}=n\sub{\dos}$ and that
the magnetization vanishes for $\ell=1,\ldots,L$.

For odd $L$, the ground state is a doublet and therefore not invariant under the
$S\sub{z}\to-S\sub{z}$ transformation: its $\ups$-spin component of doublet is transformed
into the $\dos$-spin component. Like the magnetization density under \pbc\ or \tbc, the
average magnetization in the ground-state is $1/L$. We cannot expect it to be uniform,
however, and the following analytical calculation of the magnetization density for the
$U=0$ model shows that $m\sub{\ell}$ is staggered, a conclusion that will be numerically
extended to $U\ne0$ in Sec.~\ref{sec:16}.

With $U=0$, the model Hamiltonian can be written in the diagonal form~\eqref{eq:56}. For
odd $N=L$, in order of increasing energy $\epsilon\sub{k}$, the $\ups$-spin component of
the ground state comprises $(N-1)/2$ doubly-occupied single-particle levels $\dn{k}$ and
one level with $\ups$-spin occupation. The doubly occupied levels make no contribution to
the magnetization. The magnetization is entirely due to the contribution from the lone
$\ups$-spin electron, which lies at the Fermi level. Its momentum $k\sub{F}$ is the middle
element in the sequence on the \rhs\ of Eq.~\eqref{eq:50}, i.~e.,
\begin{align}
  \label{eq:68} k\sub{F} = \dfrac{\pi}2.
\end{align}

The magnetization density $m\sub{\ell}$, which is the ground-state expectation value of
$\cd{\ell\ups}\cn{\ell\ups}-\cd{\ell\dos}\cn{\ell\dos}$, can therefore be calculated from
the expression
\begin{align}
  \label{eq:69} m\sub{\ell} = \bvac\dn{k\sub{F}\ups}
(\cd{\ell\ups}\cn{\ell\ups}-\cd{\ell\dos}\cn{\ell\dos})\ddg{k\sub{F}\ups}\kvac.
\end{align}

The expectation value of $\dn{k\sub{F}\ups}\cd{\ell\dos}\cn{\ell\dos}\ddg{k\sub{F}\ups}$
being equal to zero, Eq.~\eqref{eq:69} reduces to the expression
\begin{align}
  \label{eq:70} m\sub{\ell} =
  \{\dn{k\sub{F}\ups},\cd{\ell\ups}\}\{\cn{\ell\ups},\ddg{k\sub{F}\ups}\},
\end{align} which, according to Eq.~\eqref{eq:49}, is equivalent to the relation
\begin{align}
  \label{eq:71} m\sub{\ell} = \dfrac{2}{L+1}\sin^{2}\Big(\dfrac{\pi\ell}2\Big).
\end{align}

For $U=0$, the magnetization density is therefore $2/(L+1)$ at the odd sites and zero at
the even ones. As Appendix~\ref{sec:14} shows, Coulomb repulsion enhances the
amplitude of this staggering, without affecting its phase.

\section{Numerical results}
\label{sec:17} This section presents results for the ground-state energies, and energy
gaps for the one-dimensional half-filled Hubbard model under \obc, \pbc, and \tbc\ (global
twist $\Theta=\pi L/2$) with $L=2$\textendash$7$, and for the magnetization densities for
$L=3$ and $7$.  We have fixed the chemical potential at $\mu=-U/2$, which enforces
particle-hole symmetry, and have computed the gap for excitations from the $N=L-1$ to the
$N=L$ ground states. In all cases, we compare the energies and gaps with the Lieb-Wu
prediction for the infinite system.

To compute energies, gaps, and magnetization, we have projected the model Hamiltonian upon
a real-space basis comprising the $4^{L}$ states corresponding to the four possible
occupations ($\kvac$, $\cd{j\uparrow}\kvac$, $\cd{j\downarrow}\kvac$, and
$\cd{j\uparrow}\cd{j\downarrow}\kvac$) of each site $j$. To take advantage of the
conservation laws, we have (i) constructed a basis of states with well defined charge $N$
and $z$-component $S\sub{z}$ of the spin; (ii) diagonalized the spin operator $S^{2}$ on
that basis; and (iii), taken advantage of Bloch's Theorem (inversion symmetry) to obtain
new basis states that are eigenstates of $N$, $S^{2}$, $S\sub{z}$, and the momentum
(parity) operator $p$ ($\Pi$), for \pbc\ and \tbc\ (\obc).

Projected on the basis of the eigenstates, the model Hamiltonian reduces to a
block-diagonal matrix. Each block corresponds to a sector, labeled by $N$, $S^{2}$,
$S\sub{z}$, and $p$ or $\Pi$. Given the degeneracy among states belonging to
$2S+1$-multiplet, only the matrices for $S\sub{z}=S$ had to be diagonalized. For $L\le 7$,
the computational effort to numerically diagonalize the block matrices is relatively
small. Even for $L=9$, the computational cost is moderate: the largest matrix that must be
diagonalized has dimension 8820. As shown by the following figures, however, the results
for $L\le 7$ suffice for our discussion, so fast is the convergence to the $L\to\infty$
limit.

For $L=2$ under \tbc\ since $\tau=-\tch$, the kinetic term $-\tch\cd{1}\cn{2}+\hc$ cancels
out against the twisted term $-\tau(\cd{2}\cn{1}+\hc)$, and the two sites become
decoupled. The model Hamiltonian is then trivially diagonalized. Under \obc\ or \pbc, the
largest Hamiltonian blocks have dimension 2 and can also be analytically diagonalized.
In all other cases, the ground-state energies and gaps were computed from the numerical
diagonalization of the matrices into which the conservation laws separated the projected
Hamiltonian. The ground-state energy $E\sub{\Omega}$ is the lowest eigenvalue resulting
from all diagonalizations.

To determine the energy gap $E\sub{g}$, we have computed the difference
\begin{align}
  \label{eq:110} E\sub{g} = E_{min}^{N-1}- E\sub{\Omega},
\end{align} between the ground-state energy and the minimum energy among the sectors with
$N-1$ electrons.  An alternative gap can be computed from the difference
\begin{align}
  \label{eq:111} \tilde E\sub{g} = E_{min}^{N+1}-E\sub{\Omega},
\end{align} where $E_{min}^{N+1}$ is the minimum energy among the sectors with $N+1$
electrons.

At half filling, a particle-hole transformation takes $E_{min}^{N-1}\rightleftharpoons
E_{min}^{N+1}$ and leaves $E\sub{\Omega}$ unchanged. It follows from the invariance of the
Hamiltonian under the transformation and from Eqs.~\eqref{eq:110}~and \eqref{eq:111} that
the two gaps are identical.

\subsection{Ground-state energy}
\label{sec:24}

\begin{figure}[h!]
  \centering
  \includegraphics[width=\columnwidth]{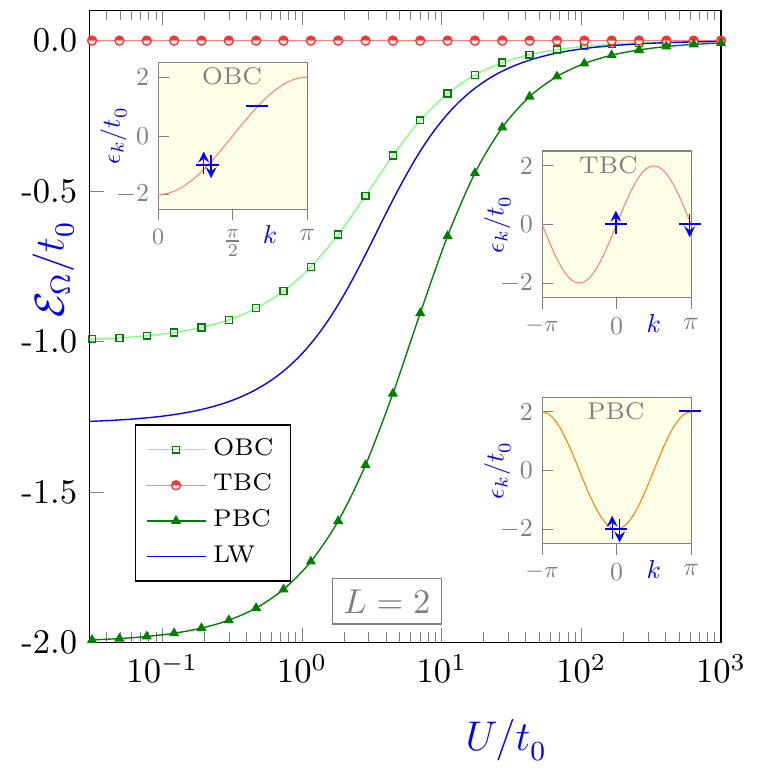}
  \caption{Per-site ground-state energies for a Hubbard dimer under periodic (triangles),
  open (open
    squares), and twisted (half-filled circles) boundary conditions as a function of the
    Coulomb parameter. The half-filled circles were computed with the special torsion
    $\Theta=(\pi/2)L=\pi$, so that the kinetic energy on the \rhs\ of Eq.~\eqref{eq:31}
    vanishes, because the term with $\ell=1$ in the sum defining the kinetic energy
    cancels the term with $\ell=2$. The blue solid line represents $N=L\to\infty$ limit,
    Eq.~\eqref{eq:101}.\cite{1968LiW1445,2003LiW1} The insets show the $L\to\infty$, $U=0$
    dispersion relation for $U=0$ under each boundary condition, the allowed levels for
    $L=2$ and their occupations for $N=2$.}
  \label{fig:5}
\end{figure}

Figure~\ref{fig:5} shows the per-site ground-state energies $\mathcal{E}\sub{\Omega}$ for
the $L=2$ model as functions of the Coulomb repulsion $U$. Under \tbc, with the special
torsion $\Theta=(\pi/2)L$, the two sites are decoupled from each other. Each site then
accommodates one electron, and the ground-state energy vanishes for all $U$. The squares
representing \obc\ and the triangles representing \pbc\ follow the trend set by the blue
solid line, which represents the Bethe-Ansatz expression~\eqref{eq:101}. The squares come
substantially closer to the $L\to\infty$ data than the triangles.

The three insets show the $U=0$, $L\to\infty$ dispersion relations under the three
boundary conditions. The bold blue dashes display the allowed levels for $L=2$, and the
arrows indicate their occupation for $N=2$. The ground-state energy is the sum of the
single-particle energies for the occupied levels, which coincides with the $U\to0$ limits
of the corresponding curves in the main plot, that is, $\mathcal{E}\sub{\Omega} =0,
-\tch$, and $-2\tch$ for \tbc, \obc, and \pbc, respectively.

Clearly, the dimer is exceptional, especially so under \tbc. We therefore turn to larger
lattices. Since, as discussed in Section~\ref{sec:6}, even- and odd-$L$ Hamiltonians
behave differently under particle-hole transformations, we will consider $L=3$, 5, and 7
first, and then $L=4$, and 6.

\begin{figure}[h!]
  \centering
  \includegraphics[height=0.78\textheight]{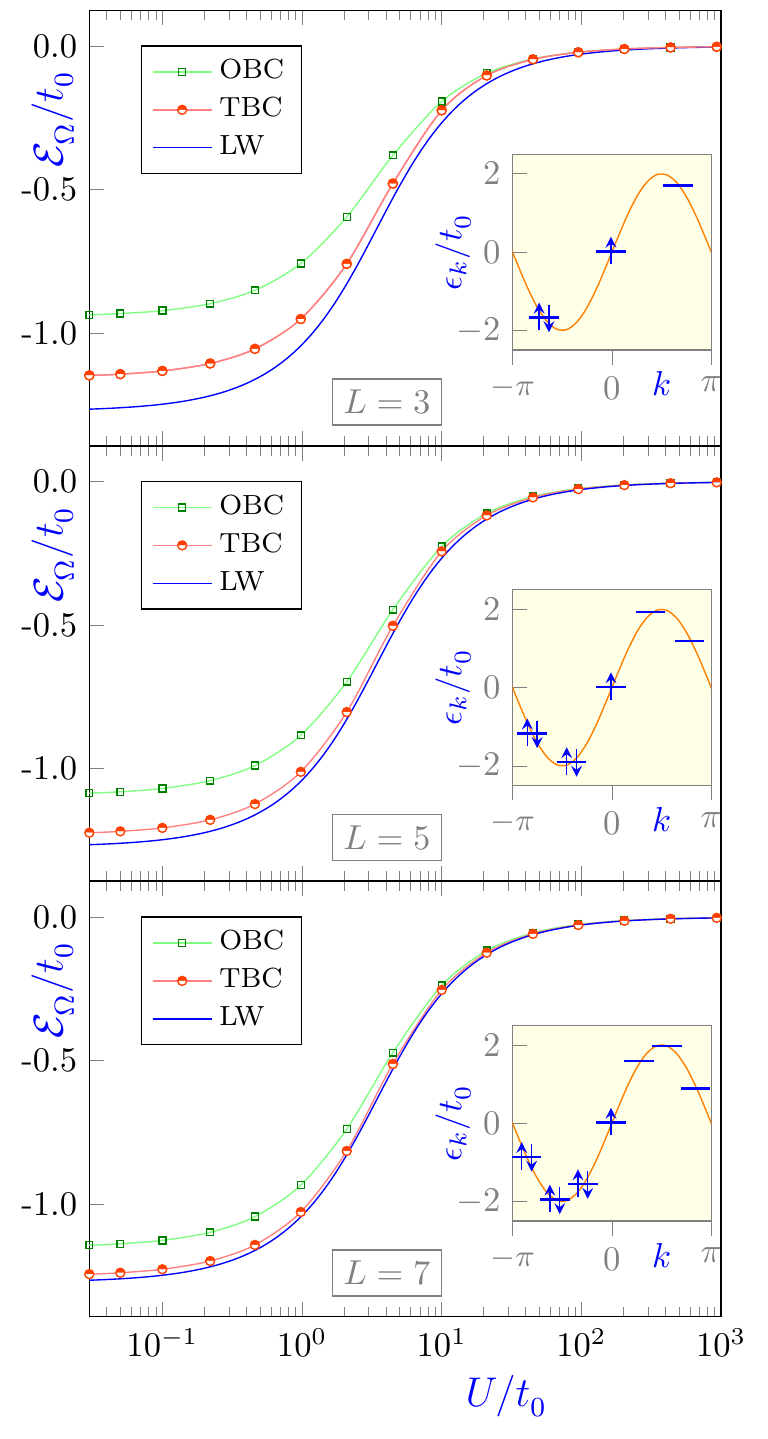}
  \caption{Per-site ground-state energies for the one-dimensional Hubbard model with $L=3$
  (top
    panel), $L=5$ (central panel), and $L=7$ (bottom panel) under twisted
    ($\Theta=L\pi/2$) and open boundary conditions as functions of Coulomb repulsion. The
    symbol convention follows that in Fig.~\ref{fig:5}. The inset shows the $L\to\infty$,
    $U=0$ dispersion relation for twisted boundary condition, the allowed levels for $L$
    sites, and their ground-state filling for $N=L$. }
\label{fig:6}
\end{figure}

Figure~\ref{fig:6} shows the per-site energies as functions of $U$ for $L=3$, 5 and
7. Particle-hole symmetry being incompatible with \pbc\ for odd $L$, only the results for
\obc\ and \tbc\ [$\Theta=(\pi/2) L$] are shown. As can be seen from the sequence of
panels, the red half-filled circles representing $\Theta=(\pi/2)L$ rapidly approach the
$L\to\infty$ limit, the disagreement with the blue solid line being substantially smaller
than the deviations between the green open squares (\obc) and the blue line.

\begin{figure}[h!]
  \centering
  \includegraphics[height=0.68\textheight]{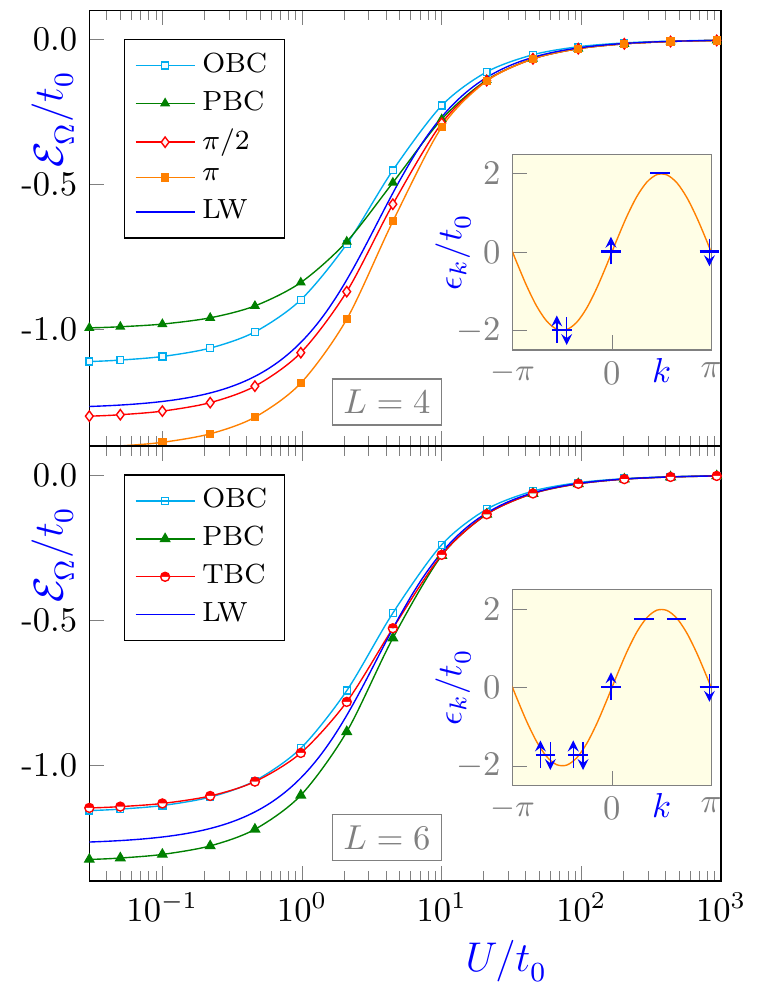}
  \caption{Per-site ground-state energies for the Hubbard Hamiltonian with $L=4$ (top
  panel), and
    $L=6$ (bottom panel) under various conditions, as functions of Coulomb repulsion. The
    blue solid line depicts the $L\to\infty$ limit, Eq.~\eqref{eq:101}. In the top panel,
    $\Theta=(\pi/2)L$ yields $\Theta=2\pi$, which is equivalent to periodic boundary
    condition, and results for two other torsions are shown: $\Theta=\pi/2$, and $\pi$.
    The bottom panel shows the ground-state energy for open, periodic, and twisted
    [$\Theta=(\pi/2)L$] boundary conditions. The top and bottom insets show the $U=0$,
    $L\to\infty$ dispersion relation for $\Theta=(\pi/2)L$, the allowed levels for $L=4$
    and $L=6$, and their ground-state fillings for $N=4$ and $N=6$, respectively.}
\label{fig:7}
\end{figure}

As suggested by the data in Fig~\ref{fig:3}, however, the convergence for even $L$ is
significantly slower. Figure~\ref{fig:7} depicts the per-particle ground-state energies
for $L=4$ (top panel), and $L=6$ (bottom panel) under \obc, and \pbc. For $L=4$ the latter
condition is equivalent to the special torsion $\Theta=(\pi/2)L=2\pi$. For comparison, the
top panel also shows results for $\Theta=\pi/2$, which conflicts with Eq.~\eqref{eq:21},
and $\Theta=\pi$. The red diamonds representing the energies for $\Theta=\pi/2$ show very
good agreement with solid black curve representing the $L\to\infty$ limit, in contrast
with the large deviations associated with \pbc.
Nevertheless, as discussed in Section~\ref{sec:25}, neither $\Theta=\pi/2$, nor
$\Theta=\pi$ yield the zero-energy single particle level at $k=0$ shown in the inset
($\Theta=2\pi$). In the absence of this level the energy gap fails to vanish as
$U\to0$. For this reason, the results for \tbc\ in the bottom panel and elsewhere in this
paper are restricted to $\Theta=(\pi/2)L$, which satisfies Eq.~\eqref{eq:21} and, for
$U=0$, positions the $k=0$ single-particle eigenvalue at $\epsilon\sub{k}=0$, as the inset
of Fig.~\ref{fig:7} shows.

\begin{figure}[h!]
   \centering
   \includegraphics[width=\columnwidth]{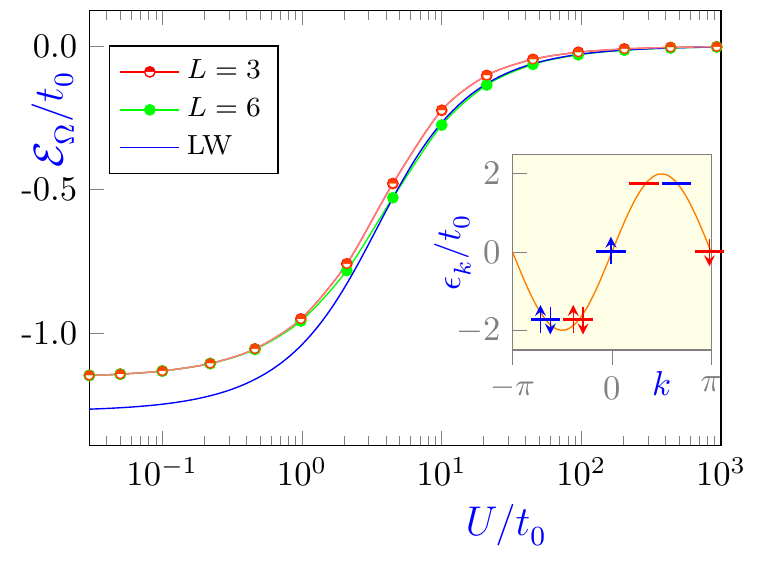}
   \caption[3vs6]{Comparison between the per-site ground-state energies for $L=3$ and
   $L=6$ under
     twisted boundary condition, with $\Theta=(\pi/2)L$. For small $U$ the red and green
     curves are virtually coincident, but the $L=6$ data approach the $L\to\infty$ limit
     faster as $U$ grows. The inset shows the three $U=0$ single-particle energy levels
     for $L=3$ (blue) and the three additional levels for $L=6$ (red).}
   \label{fig:8}
 \end{figure}

Section~\ref{sec:24} has pointed out that, under torsion $\Theta=(\pi/2)L$, the $U=0$
per-site ground-state energies for $L=2n$ ($n=1,3,\ldots$) converge relatively slowly to
the $L\to\infty$ limit because they are equivalent to the $L=n$ per-site energies. That
the equivalence is only exact for $U=0$ is shown by Fig.~\ref{fig:8}, which compares the
per-site energies for $L=3$ (half-filled circles) and $L=6$ (filled circles) as functions
of $U$. While the two curves are nearly congruent for small $U$, for larger Coulomb
repulsion the filled circles approach the $L\to\infty$ limit faster than the half-filled
circles.

 The inset explains the coincidence between the $L=3$ and $L=6$ per-site ground-state
 energies for $U=0$. The single-particle energies for $L=3$ and for $L=6$ are represented
 by bold dashes on top of the $\Theta=(\pi/2)L$ dispersion relation. Blue dashes depict
 the three $L=3$ single-particle levels, which correspond to $k=0,\pm2\pi/3$. For $L=6$,
 the allowed momenta are $k=0,\pm\pi/3, \pm2\pi/3$, and $\pi$, a sequence that can equally
 well be written as $k=0,\pm\pi/3, \pi-(\pm\pi/3)$, and $\pi-0$. In other words, to each
 $k$ in the $L=3$ sequence there correspond two momenta in the $L=6$ sequence, one with
 momentum $k$, the other with momentum $\pi-k$. It follows that the ground-state energy
 for $L=6$ is twice the one for $L=3$, and the per-site energies are identical. The same
 reasoning identifies the $U=0$ per-site ground-state energies for $L=5,7,9,\ldots$ with
 those for $L=10,14,18,\ldots$, respectively.

\subsubsection{Convergence as a function of filling}
\label{sec:36}
\begin{figure}[!h]
  \centering
  \includegraphics[width=0.95\columnwidth]{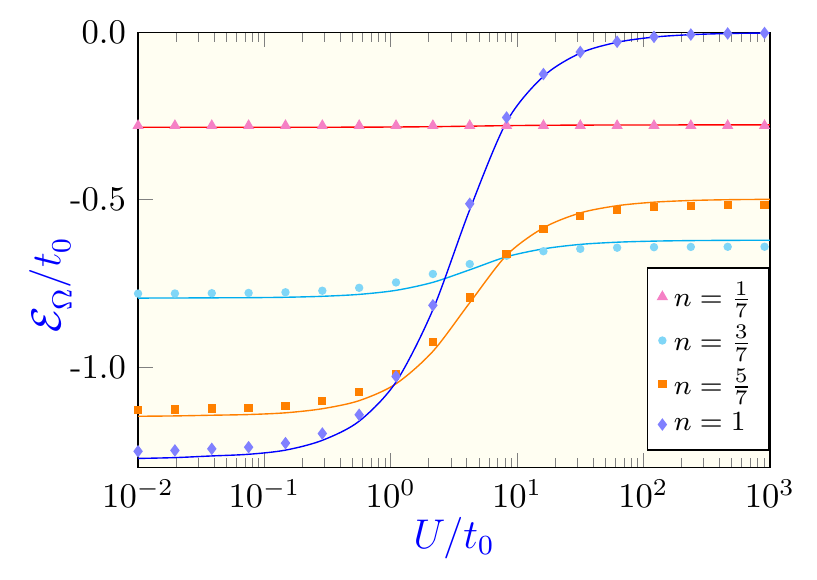}
  
  \caption{(Color online) Ground-state energies as functions of the Coulomb repulstion $U$ for the indicated
    uniform densities, under twisted boundary condition with torsion
    $\Theta=(\pi/2)L$. The solid lines are the ground-state energies resulting from the solution of 
    the Lieb-Wu equations for the infinite lattice with $n=1/7$ (magenta), $3/7$ (cyan), $5/7$
    (orange), and 1 (blue). The symbols represent the ground-state energies for lattice-size $L=7$
    with one, three, five, and seven electrons, respectively.}
  \label{fig:13}
\end{figure}

Figure~\ref{fig:13} shows the ground-state energies calculated under twisted boundary condition
with the special torsion $\Theta=(\pi/2)L$ for $L=7$ with one (magenta triangles), three
(cyan circles), five (orange squares), and seven (blue diamonds) electrons. For comparison, the
ground-state energies for the infinite lattice with the same uniform electron densities are shown by
the solid lines of the same colors. Given particle-hole symmetry, we need not display results
between $n=1$ and $n=2$, since $\mathcal{E}\sub{\Omega}(n)=\mathcal{E}\sub{\Omega}(2-n)$. 
 
For the intermediate densities $n=3/7$ and $n=5/7$, the numerical results at large $U/\tch$ can be
seen to slightly underestimate the infinite-size model, in contrast with the very good agreements
for $n=1$ and $n=1/7$. At small $U$ the finite-size energies slightly overestimate those of the
infinite system at every density. In all cases, however, the $L=7$ energies represent the infinite
limit well, with less than 5\% deviations. Although our discussion in other sections is limited to
half filling, the conclusions are general.

\subsection{Energy gap}
\label{sec:25}

\begin{figure}[h!]
  \centering
  \includegraphics[height=0.88\textheight]{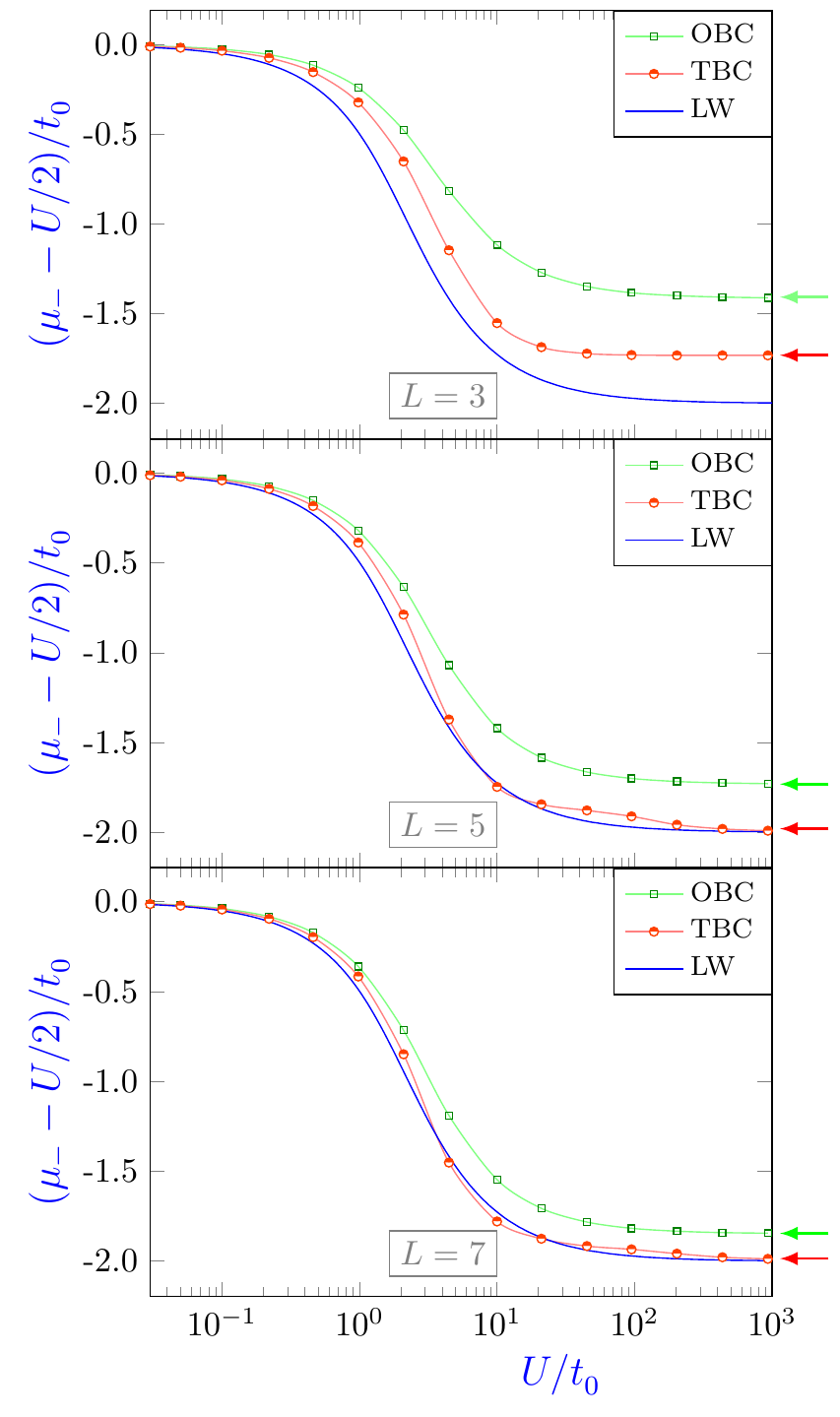}
  \caption[]{Energy gaps as functions of Coulomb energy for $L=3,5,7$. The gap is always
  measured
    from the chemical potential $\mu=U/2$ so that the plot approaches a finite limit as
    $U\to\infty$.}
\label{fig:9}
\end{figure}

\begin{figure}[h!]
  \centering
  \includegraphics[width=0.9\columnwidth]{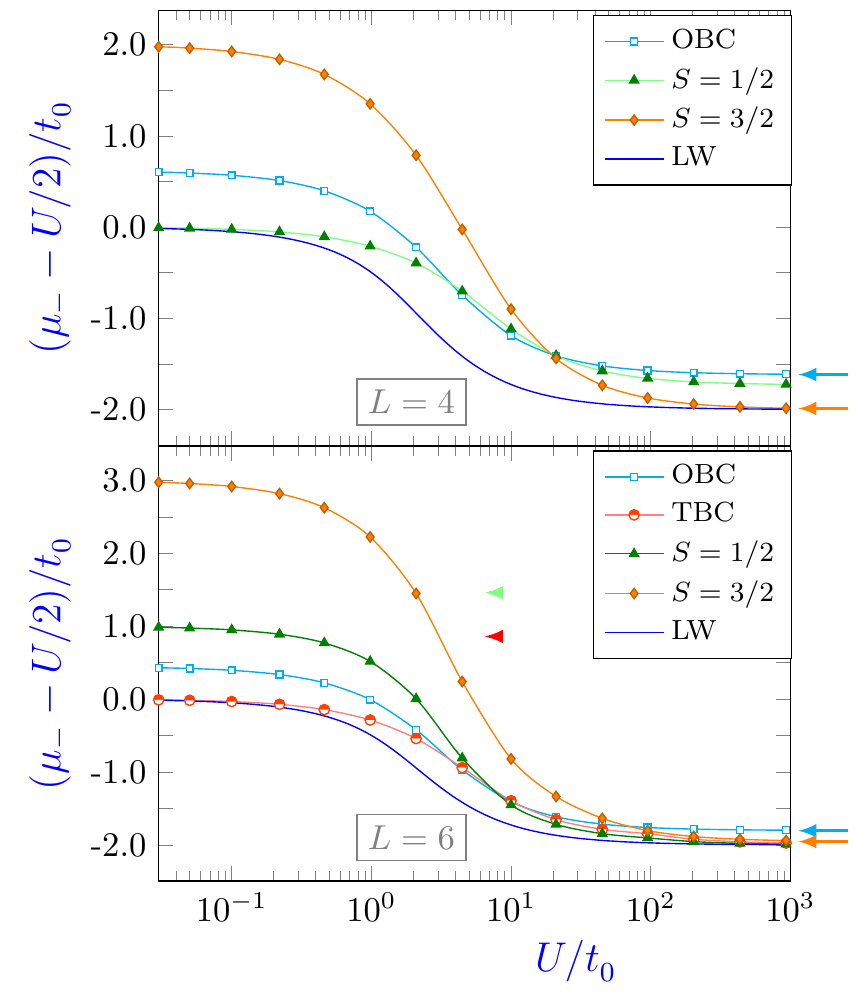}
  \caption[]{Energy gap as a function of Coulomb energy for $L=4$ and $6$, top and bottom
  panels,
    respectively. The gaps are measured from the chemical potential $\mu=U/2$ to insure
    convergence to a finite limit as $U\to\infty$. In both panels, the open squares
    represent open boundary condition. The triangles and diamonds represent the gaps
    measured from the lowest energies in the sectors with spin $S=1/2$ and $S=3/2$,
    respectively under twisted boundary condition with the special torsion $\Theta=(\pi
    L)/2$. The solid curve represents the gap in the $L\to\infty$ limit.}
\label{fig:10}
\end{figure}

Figure~\ref{fig:9} displays the energy gaps for $L=3,5$ and 7 as functions of the Coulomb
repulsion, for \obc\ and \tbc\ [$\Theta=(\pi/2)L$]. The gaps are measured from the
chemical potential, so that they approach a finite limit, $\mu_{-}^{\infty}-U/2=-2\tch$,
as $U,L\to\infty$.  For $U\to\infty$ with finite $L$, the horizontal arrows pointing to
the right-hand vertical axes indicate the gaps expected from Eqs.~\eqref{eq:72}~and
\eqref{eq:74}, under \obc, or from Table~\ref{tab:3}, under \tbc.

For small $U$, the open squares, which represent \obc, lie close to the solid line
representing the Lieb-Wu result.\cite{1968LiW1445,2003LiW1} The deviations between the
squares and the continous line grow with $U$ and monotonically approach the $U\to\infty$
limits. The red half-filled circles, which represent \tbc, show similar behavior, but two
distinctions are noteworthy: (i) only for $L=3$ there is significant vertical separations
between the red arrows and the $U,L\to\infty$ limit; and (ii) in all panels, the half-open
circles approach the solid line much faster than the open squares.

The apparent oscillations and plateaus in the red curves reflect the $U$ dependence of the
ground-state spin $S$. For $L=5$, $N=4$, for instance, the ground-state is a singlet for
$U=0$, but $S$ evolves as $U$ grows and imposes an increasing penalty on double
occupation. Let $E\sub{S}$ denote the minimum energy in the sector with spin $S$. Relative
to $E\sub1$, the energy $E\sub0$ grows with $U$ until it exceeds $E_{1}$, at which point
the ground state shifts from the $S=0$ to the $S=1$ sector. Table~\ref{tab:3} confirms
that, in the $U\to\infty$ limit, the ground state has spin $S=1$.

More explicit information is provided by Fig.~\ref{fig:10}, which show the energy gaps as
functions of Coulomb repulsion for $L=4$ and 6. The triangles and diamonds represent the
gaps under \tbc, computed as the differences between the lowest energies in the $N=L-1$
sectors with $S=1/2$ and $S=3/2$, respectively, and the ground-state energy for $N=L$. For
small $U$, in both panels, the lowest energy in the $S=1/2$ sector is smaller than in the
$S=3/2$ sector and hence yields the smaller gap. As $U/\tch$ grows, however, the curve
through the diamonds drops faster than the curve through the triangles. In the top panel,
the two curves cross around $U=20\tch$. For $U>20\tch$, the lower energy lies in the
sector with $S=3/2$. The energy gap under \tbc\ therefore follows the triangles from $U=0$
to $U\approx20\tch$ and the diamonds for $U>20\tch$. In the bottom panel, the lowest
energies in the two sectors become degenerate in the $U\to\infty$ limit, and the energy
gap is described by the triangles for any Coulomb repulsion. The $U\to\infty$ limits of
both panels agree with the results in Table~\ref{tab:3}, which show that the $N=L-1$
ground state has spin $S=3/2$ for $L=4$ and spins $S=1/2$ or $3/2$ for $L=6$, and yield
the gaps indicated by the orange horizontal arrows pointing to the right-hand vertical
axes.

Under \obc, for $U=0$, the single-particle spectra contain no zero energy, as a result of
which a gap of the order of $1/L$ opens, in disagreement with the zero gap predicted by
the Bethe-Ansatz solution.\cite{1968LiW1445,2003LiW1} The open squares representing \obc\
in Fig.~\ref{fig:10} show similar discrepancies for all Coulomb repulsions. Compared with
the results under \tbc, the plots in the two panels show inferior agreement with the
$L\to\infty$ limit both for $U\ll\tch$ and $U\gg\tch$. Only for intermediate Coulomb
repulsions are the deviations between the gaps under \obc\ comparable to those computed
under \tbc.

\subsection{Magnetization density}
\label{sec:16}

As explained in Sec.~\ref{sec:22}, at half filling the electronic density is uniform,
$n\sub{\ell}=1$ ($\ell=1,\ldots,L$), under \obc, \pbc, or \tbc. The magnetization density
vanishes identically under \obc, \pbc, or \tbc\ for even $N=L$. For odd $N=L$, it is
uniform under \pbc\ and \tbc: $m\sub{\ell}=1/L$ ($\ell=1,\ldots,L$).
For odd $N=L$ under \obc\ we have found the $U=0$ magnetization density to be
staggered. Here we present numerical results for $U\ne0$. 

\begin{figure}[h!]
  \centering
  \includegraphics[width=\columnwidth]{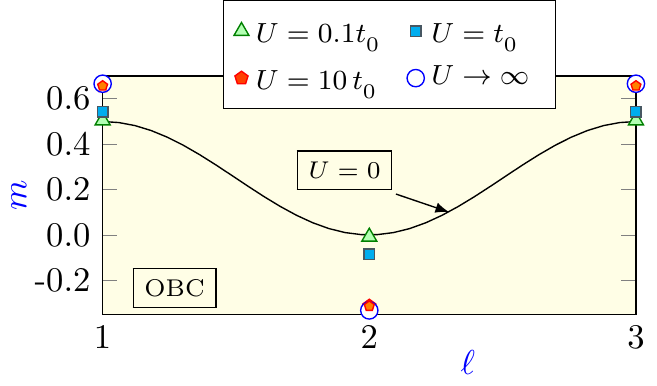}
  \caption{Magnetization density as a function of lattice position for the Hubbard trimer
    under open boundary condition. The solid black line represents Eq.~\eqref{eq:71}. The
    filled triangles, filled squares, and filled circles were obtained via numerical
    diagonalization of the model Hamiltonian, for the indicated Coulomb repulsions
    $U$. The open circles represent the $U\to\infty$ limits obtained in Sec..
}
  \label{fig:11}
\end{figure}

Figure~\ref{fig:11} plots the magnetization density as a function of site position for the
half-filled Hubbard trimer under \obc. With $L=3$, the $U\to\infty$ magnetization density is given
by Eq.~\eqref{eq:109}, which is depicted by open circles. The filled triangles, squares, and circles
show that the magnetization density at the borders ($\ell=1,3$) progressively rises from
$m\sub{\ell}=1/2$ to $m\sub{\ell}=2/3$ as $U/\tch$ grows. At the center ($\ell=2$) the magnetization
density becomes negative and likewise progresses towards the $U\to\infty$ limit ($m\sub2=-1/3$).

\begin{figure}[h!]
  \centering
  \includegraphics[width=\columnwidth]{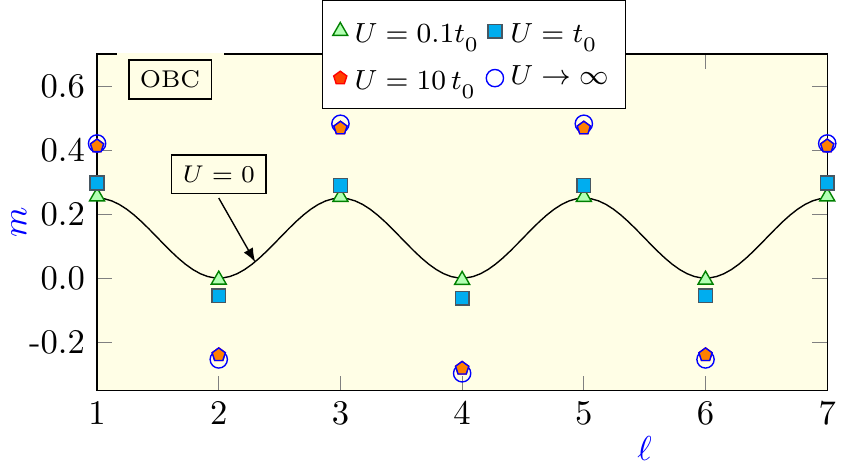}
  \caption[mag]{Magnetization density for the seven-site Hubbard model with the indicated
  Coulomb
    repulsions. The solid line represents the analytical expression for $U=0$,
    Eq.~\eqref{eq:71}. All other data were calculated numerically. The open circles were
    obtained from the ground state of the $U=250\tch$ model and cannot be distinguished,
    on the scale of the figure from the magnetization densities computed for larger
    $U/\tch$.}
  \label{fig:12}
\end{figure}

For longer lattices with odd $L$, the evolution of the magnetization density as $U$ grows is
similar, as illustrated by Fig.~\ref{fig:12}. The staggered pattern in Fig.~\ref{fig:11} is
reproduced. In particular, the amplitude of the oscillations is enhanced as $U$ grows and the
magnetization becomes negative at the even-$\ell$ sites. The enhancement is more pronounced in the
central region than near the borders. Inspection of our results for different lattice sizes has
shown that the amplitude of the oscillation is of $\mathcal{O}(1/L)$. The magnetization density
therefore vanishes uniformly as $L\to\infty$.

\section{Conclusions}
\label{sec:26}
In this paper, we have focused on the effect of boundary conditions on small systems, which are of
increasing importance with the progressive shrinking and control of nanoscale systems. Our results
could be already of relevance for recent experiment on, e.~g., Bose-Einstein
condensates\cite{2016Cheuk1260,2016Boll1257,2016Parsons1253} or ultracold fermionic
atoms.\cite{2015Murmann080402} We have examined the finite-size one-dimensional Hubbard model and
compared two of its ground-state properties, the ground-state energy and the energy gap, with those
of the infinite system. We have concentrated our attention on the energy of the half-filled and
nearly half-filled one-dimensional models because the corresponding eigenvalues of the
infinite-lattice model have been exactly computed by Lieb and Wu, allowing meaningful comparisons.

The chosen model Hamiltonian is also convenient because it remains invariant under a number of
symmetry operations, which have served as beacons in our analysis. Not all boundary
conditions preserve the symmetry of the infinite lattice. Open boundary condition, for
instance, is inconsistent with translational invariance, and \pbc\ only preserves
particle-hole when the number $L$ of lattice sites is even. The numerical results in
Sec.~\ref{sec:17} have shown that the $L$ dependent ground-state energy and gap can
display rapid or slow convergence to the infinite limit, depending on whether the
symmetries are or not preserved.

Chiefly important, in this context, is torsion. As Sec.~\ref{sec:6} has shown, \tbc\
preserves translational symmetry. The special torsion $\Theta=(\pi/2)L\mod \pi$ also
preserves particle-hole symmetry. Left-right inversion symmetry is only preserved for
$\Theta=0\mod\pi$, which is inconsistent with the special torsion for odd
$L$. Left-right asymmetry has no effect upon the computed properties, however, since
inversion amounts to relabeling the momenta, $k\to-k$.  Overall, small $L$ models under
\tbc\ with $\Theta=(\pi/2)L\mod \pi$ offer the most faithful representation for the
properties of the infinite model. As illustrated by the diamonds in the top panel of
Fig.~\ref{fig:7}, the torsion can be adjusted to yield nearly perfect agreement with the
ground-state energy of the infinite model; the adjustment nonetheless breaks particle-hole
symmetry and hence yields poor agreement for the energy gap.

The ground-state energy is sensitive to translational invariance, and the energy gap to
particle-hole symmetry. Neither is preserved under \obc, which hence yields relatively
slow convergence to the infinite-lattice limit. Under \pbc, translational symmetry is
always preserved, but the odd-$L$ models are particle-hole asymmetric. It results that,
for odd $L$, the gap deviations from the $L\to\infty$ limit under \pbc\ are comparable to
those under \obc. Under \tbc\ with $\Theta=(\pi/2)L\mod\pi$, both the ground-state energy
and the gap for finite-size models rapidly approach the $L\to\infty$ limit.

Twisted boundary condition has proved instrumental in numerical analyses of finite-size models
targeting the thermodynamical limit. We have shown that the symmetry-preserving torsion
$\Theta=(\pi/2)L\mod\pi$ insures rapid convergence and may hence be especially valuable in studies
of models that remain invariant under particle-hole transformation.

\acknowledgments KZ and LNO gratefully acknowledge financial support from the FAPESP
(Fellowship grant no.~12/02702-0), CNPq (grants no.~312658/2013-3 and ~140703/2014-4) and CAPES (Scholarship grant no.~88881.135185/2016-01).  ID likewise
acknowledges support from the Royal Society through the Newton Advanced Fellowship scheme
(grant no.~NA140436). Finally, this work would not have been possible without a PVE grant
(no.~401414/2014-0) from the CNPq.

\appendix

\section{\label{sec:29} Analytical results for $U=0$}

\subsection{\label{sec:11} Closed boundary conditions}

For \pbc\ or \tbc, Bloch's Theorem associates each single-particle eigenstate of
$\textbf{H}$ with a unique momentum $k$. Since the number of basis states $\ad{\ell}$ is
$L$, we will have to define $L$ distinct momenta. For now, however, we let the $k$'s be
undetermined parameters.

Since $\mathbf{H}$ remains invariant under lattice translations, the model Hamiltonian
commutes with the unit-translation operator $T\sub{1}$, defined by the identity
\begin{align}
  \label{eq:32} T\sub{1}\ad{\ell}\kvac = \ad{\ell+1}\kvac.
\end{align} with the operators $\an{\ell}$ defined by Eq.~\eqref{eq:29}.

We seek eigenvectors of $T\sub{1}$. Promising candidates are defined by the normalized
Fermi operator
\begin{align}
  \label{eq:33} \bd{k} = \dfrac1{\sqrt L}\sum_{\ell=1}^{L}e^{-ik\ell}\ad{\ell}.
\end{align}

To verify that the $\bd{k}$ diagonalize $T\sub{1}$, we only have to compute
$T\sub{1}\bd{k}\kvac$. From Eqs.~\eqref{eq:32}~and~\eqref{eq:33} we can see that
\begin{align}
  \label{eq:34} T\sub{1}\bd{k}\kvac = \dfrac1{\sqrt
  L}\Big(\sum_{\ell=1}^{L-1}e^{-ik\ell}\ad{\ell+1}+e^{-ikL}\ad{1}\Big)\kvac,
\end{align} where we have separated the last term from the sum on the \rhs\ to emphasize
that, under \pbc\ or \tbc, the translation displaces $\ad{L}$ to $\ad{1}$, as prescribed
by Eq.~\eqref{eq:27}.

We then change the summation index to $\ell'=\ell+1$ in the sum on the \rhs\ of
Eq.~\eqref{eq:34}, which shows that
\begin{align}
  \label{eq:35}
    T\sub{1}\bd{k}\kvac = \dfrac1{\sqrt
    L}\Big(\sum_{\ell'=2}^{L}e^{-ik(\ell'-1)}\ad{\ell'}+e^{-ikL}\ad{1}\Big)\kvac.
\end{align}

To include the last term within parentheses in the sum on the \rhs, we now impose the
condition
\begin{align}
  \label{eq:36} e^{ikL}=1,
\end{align} so that Eq.~\eqref{eq:35} reduces to the compact expression
\begin{align}
  \label{eq:37}
    T\sub{1}\bd{k}\kvac = \dfrac1{\sqrt
    L}\sum_{\ell'=1}^{L}e^{-ik(\ell'-1)}\ad{\ell'}\kvac,
\end{align} which shows that
\begin{align}
  \label{eq:38}
    T\sub{1}\bd{k}\kvac = e^{ik}\bd{k}\kvac.
\end{align}

From Eq.~\eqref{eq:38} we can see that, for momenta satisfying Eq.~\eqref{eq:36}, the $\bd{k}$
are eigenstates of the translation operator. Equation~\eqref{eq:36} is equivalent to the
expression
\begin{align}
  \label{eq:39} k = \dfrac{2n\pi}{L},
\end{align} where $n$ is an integer.

To generate $L$ distinct eigenstates, we could let $n$ run from unity to $L$ on the \rhs\
of Eq.~\eqref{eq:39}. It is nonetheless customary to choose the integers so that the
momenta lie in the first Brillouin Zone, i.~e., for $-\pi<k\le\pi$. The following
sequences are therefore defined:
\begin{align}
  \label{eq:40} n=
  \begin{cases}
-\dfrac{L}2+ 1, \ldots,\dfrac{L}2\qquad&(L=\mbox{even})\\[2mm]
 -\dfrac{L-1}2, \ldots,\dfrac{L-1}2\qquad&(L=\mbox{odd}).
  \end{cases}
\end{align}

Equations~\eqref{eq:33}, \eqref{eq:39}, and \eqref{eq:40} define a set of $L$
non-degenerate eigenstates of the translation operator $T\sub{1}$. Since the latter
commutes with the Hubbard Hamiltonian $\mathbf{H}$ under \pbc\ or \tbc, we can see that
the $\bd{k}$ also diagonalize $\mathbf{H}$.

To complete the diagonalization, we have to find the eigenvalues associated with the
$\bd{k}$. On the basis of the latter, the Hubbard Hamiltonian takes the form
\begin{align}
  \label{eq:41} \mathbf{H} =\sum_{k}(\epsilon\sub{k}-\mu)\bd{k}\bn{k},
\end{align} from which we have that
\begin{align}
  \label{eq:42} [\mathbf{H},\bd{k}] = (\epsilon\sub{k}-\mu)\bd{k}.
\end{align}

To identify the eigenvalues $\epsilon\sub{k}$ we therefore need to compute the commutator
on the left-hand side of Eq.~\eqref{eq:42} and start out by computing the commutator
$[\mathbf{H},\ad{\ell}]$ between the Hamiltonian and a local operator $\ad{\ell}$
($\ell=1,\ldots,L$). From Eq.~\eqref{eq:31}, with $U=0$, we have that
\begin{align}
  \label{eq:43} [\mathbf{H},\ad{\ell}]=
  -\tun\ad{\ell+1}-\tun^{*}\ad{\ell-1}-\mu\ad{\ell}\qquad(\ell=1,\ldots,L).
\end{align}

Reference to Eq.~\eqref{eq:33} now shows that
\begin{align}
  \label{eq:44} \sqrt{L}[\mathbf{H},\bd{k}] =&
  -t\sum_{\ell=1}^{L}\Big(e^{-ik\ell}\ad{\ell+1}\Big)
-t^{*}\sum_{\ell=1}^{L}\Big(e^{-ik\ell}\ad{\ell-1}\Big)-\mu
-\sum_{\ell=1}^{L}e^{-ik\ell}\ad{\ell}.
\end{align}

We then let $\ell\to\ell-1$ in the first sum on the right-hand side and $\ell\to\ell+1$ in
the second sum. The limits of the first and second sums will change. Nonetheless, thanks
to boundary condition, which makes $\ell=0$ ($\ell=N+1$) equivalent to $\ell=N$
($\ell=1$), the sums will still cover all lattice sites, $\ell=1,\ldots,L$. It therefore
follows that
\begin{align}
  \label{eq:45} [\mathbf{H},\bd{k}] =& -te^{ik}\bd{k}-t^{*}e^{-ik}\bd{k}-\mu \bd{k}.
\end{align}

We next recall that $t\equiv\tch e^{i\theta}$, and compare with Eq.~\eqref{eq:42} to see
that
\begin{align}
  \label{eq:46} \epsilon\sub{k} = -2\tch\cos(k+\theta).
\end{align}

In particular, under \pbc\ ($\theta=0$) Eq.~\eqref{eq:46} reduces to the equality
\begin{align}
  \label{eq:47} \epsilon\sub{k} = -2\tch\cos(k),
\end{align} and under \tbc\ with the special torsion $\Theta=(\pi/2)L$, to the equality
\begin{align}
  \label{eq:48} \epsilon\sub{k} = 2\tch\sin(k).
\end{align}

\subsection{\label{sec:12} Open boundary condition}

Open boundary condition invalidates Bloch's Theorem. Instead of a running wave, we may
visualize a wave-function that vanishes at $\ell=0$ and $\ell=L+1$, an image that
associates the following single-particle operator with the single-particle eigenvectors:
\begin{align}
  \label{eq:49} \ddg{k} = \sqrt{\dfrac2{L+1}}\sum_{\ell=1}^{L}\sin(k\ell)\cd{\ell},
\end{align} subject to the condition that $\sin(k\ell)$ vanish for $\ell=L+1$, i.~e., for
momenta given by the equality
\begin{align}
  \label{eq:50} k = \dfrac{\ell\pi}{L+1}\qquad(\ell=1,\ldots,L).
\end{align}

To show that the $\ddg{k}$ in Eq.~\eqref{eq:49} diagonalize Eq.~\eqref{eq:1}, we again
compute the commutator $[\mathbf{H},\cd{\ell}]$. Under \obc\ we find that
\begin{align}
    [\mathbf{H},\cd{\ell}]&= -\mu\cd{\ell}
    -\tch\cd{2}\quad&(\ell=1);\nonumber\\
  \label{eq:51}
    [\mathbf{H},\cd{\ell}]&= -\mu\cd{\ell}
    -\tch\cd{\ell+1}-\tch\cd{\ell-1}\quad&(1<\ell<L);\\
    [\mathbf{H},\cd{\ell}]&= -\mu\cd{\ell}
    -\tch\cd{L-1}\quad&(\ell=L).\nonumber
\end{align}

From Eqs.~\eqref{eq:49}~and \eqref{eq:51} we then have that
\begin{align}
  \label{eq:52} [\mathbf{H},\ddg{k}] =
  -\sqrt{\dfrac2{L+1}}\tch\Bigg(&\sum_{\ell=2}^{L}\sin\Big(k(\ell-1)\Big)\cd{\ell}
                                  +\sum_{\ell=1}^{L-1}\sin\Big(k(\ell+1)\Big)\cd{\ell}\Bigg)-\mu\ddg{k}.
\end{align}

Since $\sin(k\ell)$ vanishes for $\ell=0$, we can let the summation index in the first sum
on the \rhs\ of Eq.~\eqref{eq:52} run from $\ell=1$ to $\ell=L$. Likewise, given that
$\sin[k(L+1)]=0$ [see Eq.~\eqref{eq:50}], we can extend the second sum to $\ell=L$, to
obtain the expression
\begin{align}
  [\mathbf{H},\ddg{k}] =
-\sqrt{\dfrac2{L+1}}\tch\sum_{\ell=1}^{L}\Bigg(&\sin\Big(k(\ell-1)\Big)
+\sin\Big(k(\ell+1)\Big)\Bigg)\cd{\ell}-\mu\ddg{k}.\label{eq:53}
\end{align}

Expansion of the sines in the summand on the \rhs\ reduces Eq.~\eqref{eq:53} to the form
\begin{align}
  \label{eq:54} [\mathbf{H},\ddg{k}] =&
  -2\tch\cos(k)\sqrt{\dfrac2{L+1}}\sum_{\ell=1}^{L}\sin(k\ell)\cd{\ell}-\mu\ddg{k}.
\end{align}

Comparison with Eq.~\eqref{eq:49} then shows that
\begin{align}
  \label{eq:55} [\mathbf{H},\ddg{k}] =& \Big(-2\tch\cos(k)-\mu\Big)\ddg{k},
\end{align} which allows us to write the \obc\ Hamiltonian in a diagonal form akin to
Eq.~\eqref{eq:41}:
\begin{align}
  \label{eq:56} \mathbf{H} =\sum_{k}(\epsilon\sub{k}-\mu)\ddg{k}\dn{k},
\end{align} with the $\epsilon\sub{k}$ from Eq.~\eqref{eq:47}.

Equation~\eqref{eq:47} describes the dispersion relations for both \obc\ and
\pbc. Nonetheless, the single-particle energies $\epsilon\sub{k}$ for \obc\ are distinct
from the $\epsilon\sub{k}$ for \pbc, because the allowed momenta are boundary-condition
dependent. For \obc, the $k$ are given by Eq.~\eqref{eq:50}; for \pbc, they are determined
by Eqs.~\eqref{eq:39}~and \eqref{eq:40}.

\subsection{\label{sec:13} Dispersion relations}

\begin{figure}[h!]
  \centering
  \includegraphics[height=0.64\textheight]{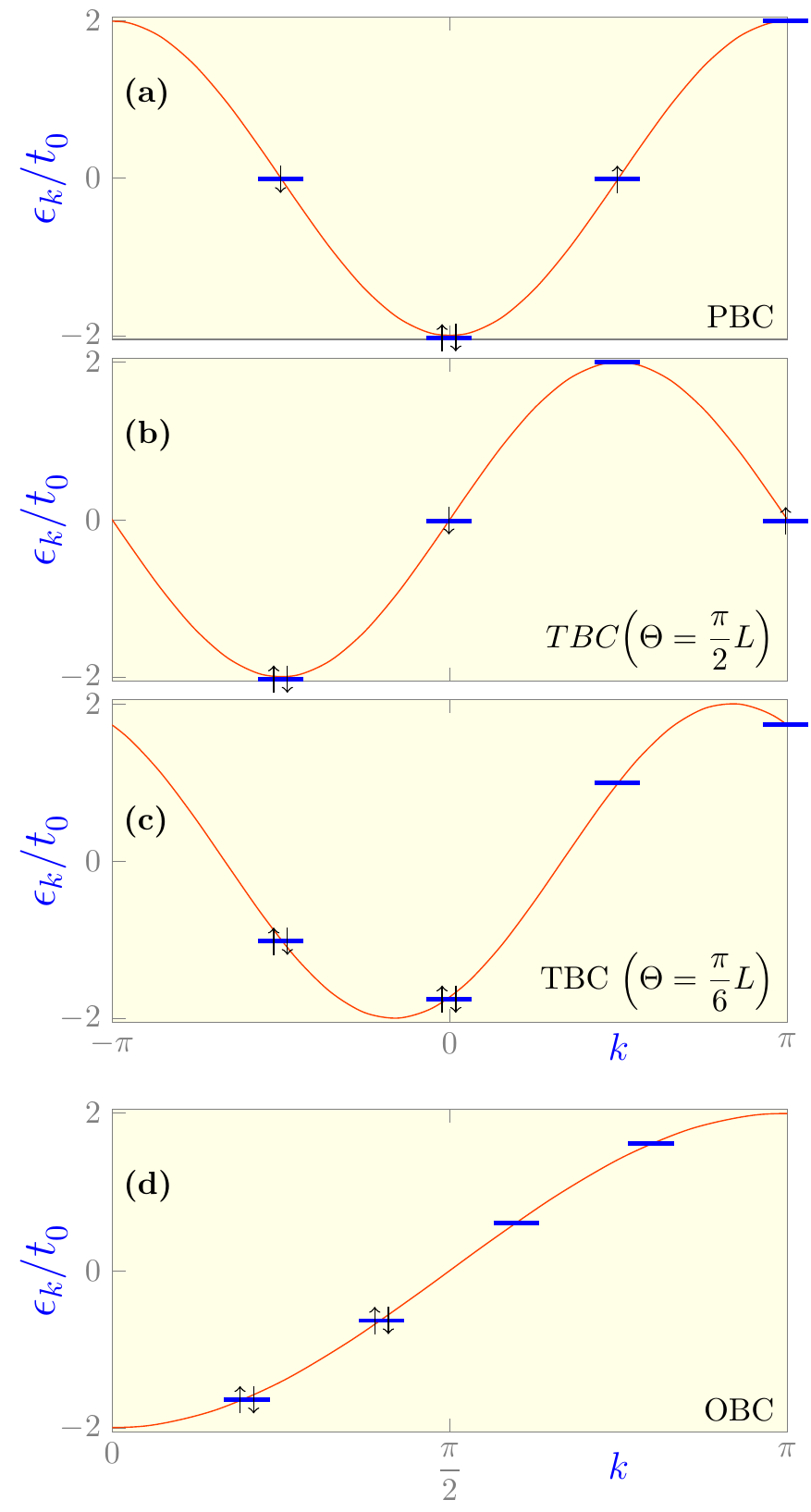}
  \caption[Dispersion relations]{Dispersion relations for (a) periodic boundary condition, (b)
    twisted boundary condition with torsion $\Theta=(\pi/2)L$ (local torsion
    $\theta=\pi/2$), (c) twisted boundary condition with torsion $\Theta=(\pi/6)L$ (local
    torsion $\theta=\pi/6$), and (d) open boundary condition. In each plot bold blue
    dashes indicate the single-particle levels for $L=4$. At half filling, the chemical
    potential is $\mu=0$, the negative-energy levels are doubly occupied, the zero-energy
    levels are singly occupied, and the positive-energy levels are vacant, as indicated by
    the vertical arrows. The dispersion relation is an even function of $k$ for periodic
    boundary condition, and an odd function for twisted boundary condition with the
    special torsion $\Theta=(\pi/2)L$. By contrast, for $\Theta=(\pi/6)L$ the dispersion
    relation is asymmetric. }

  \label{fig:2}
\end{figure}

Figure~\ref{fig:2} compares the dispersion relations for \pbc, \tbc, and \obc. As an
illustration, the single-particle levels for $L=4$ are depicted for each condition. The
single-particle levels for \pbc\ or \obc\ are given by Eqs.~\eqref{eq:47} with $k$ defined
by Eqs.~\eqref{eq:39}~or \eqref{eq:50}, respectively. Under \tbc\ the levels are given by
\eqref{eq:46}, with $k$ defined by Eq.~\eqref{eq:39}. With $\mu=0$, which corresponds to
ground-state occupation $N=4$, the levels with $\epsilon\sub{k}<0$ are doubly occupied in
the ground state, while the levels at $\epsilon\sub{k}=0$ have single occupation.

With $L=4$, the special torsion in Eq.~\eqref{eq:22} is $\Theta=2\pi$, equivalent to
$\Theta=0$. The energy levels for \pbc\ and for \tbc\ must therefore be
identical. Comparison between panels (a) and (b) in the figure shows how two distinct sets
of allowed moment can yield the same single-particle energies. Under \tbc\ with
$\Theta\ne2\pi$ [panel~(c) in Fig.~\ref{fig:2}] or \obc\ [panel~(d)] the single-particle
energies are different; there is no zero-energy level, for instance.

The single-particle spectra in panels (a), (b), and (d) of Fig.~\ref{fig:2} are
particle-hole symmetric. The bold dashes occur in pairs with energies $\pm\epsilon$, even
though their momenta are changed under particle-hole transformation: for positive $k$, for
instance, $k\to \pi-k$ in panels (a), (b), and (d). The dispersion relation in panel (b),
\tbc\ with torsion $\Theta=(\pi/2) L$, is an odd function of $k$, a symmetry that, for all
$L$, introduces a zero-energy level, at $k=0$ in the single-particle energy spectrum.  The
special torsion $\Theta=(\pi/2) L$ therefore reproduces the feature of the
infinite-lattice $U=0$ model responsible for the vanishing energy gap at half-filling. No
such zero-energy level is found in panel~(c) of Fig.~\ref{fig:2}, which is particle-hole
asymmetric, like all spectra for \tbc\ with $\Theta\ne(\pi/2)L$.

Depending on boundary condition, the $U=0$ infinite-lattice Hamiltonian can have any of
the dispersion relations represented by red solid lines in Fig.~\ref{fig:2}. With
$L\to\infty$, all momenta in the range $-\pi<k\le\pi$ are allowed, and at least one of
them will satisfy $\epsilon\sub{k}=0$. Under \obc, for example, the single-particle energy
vanishes at $k=\pi/2$. If $N=L$, at zero temperature all levels below (above)
$\epsilon\sub{k}=0$ will be filled (vacant), and the zero-energy level guarantees that it
will cost zero energy to add or to remove an electron from the ground state. There is no
energy gap.

\subsection{\label{sec:5} Ground-state energy}

In the ground state, all levels below the Fermi level are filled. If we introduce the
notation $\kocc$ to denote the momenta of the occupied levels, the expression for the
ground-state energy under \pbc\ or \obc\ reads
\begin{align}
  \label{eq:57} E\sub{\Omega}=-4\tch \sum_{\kocc}\cos(k),
\end{align} where the momenta are specified by Eqs.~\eqref{eq:39}~or \eqref{eq:50},
respectively, and the single-particle energies from~\eqref{eq:47} have been doubled to
account for spin degeneracy.

Under \tbc\ the momenta are again given by Eq.~\eqref{eq:39}, and the single-particle
energies, from~\eqref{eq:46}, which yields
\begin{align}
  \label{eq:58} E\sub{\Omega}=-4\tch \sum_{\kocc}\cos(k+\theta),
\end{align} which for the special torsion $\Theta\equiv L\theta= (\pi/2)L$ reduces to the
form
\begin{align}
  \label{eq:59} E\sub{\Omega}=4\tch \sum_{\kocc}\sin(k).
\end{align}

For all $L$, the ground-state energy can always be analytically computed, but the computation depends
on boundary condition and $L$ parity. For \obc\ and even $L$, for instance,
Eq.~\eqref{eq:57} reads
\begin{align}
  \label{eq:60} E\sub{\Omega}=-4\tch
  \sum_{\ell=1}^{L/2}\cos\Big(\dfrac{\pi\ell}{L+1}\Big).
\end{align}

It proves convenient to rewrite the right-hand side of Eq.~\eqref{eq:60} as the real part
of a complex number:
\begin{align}
  \label{eq:61}
    E\sub{\Omega}=-4\tch \Re\sum_{\ell=1}^{L/2}\exp(\dfrac{i\pi\ell}{L+1}),
\end{align} because the summand then defines a geometric progression, which can be easily
summed. The following expression results:
\begin{align}
  \label{eq:62}
    E\sub{\Omega}=-4\tch
    \Re\dfrac{i\exp\Big(\dfrac{i\pi}{2(L+1)}\Big)-\exp(\dfrac{i\pi}{L+1})}{\exp(\dfrac{i\pi}{L+1})-1}.
\end{align}

We then multiply the fraction on the \rhs\ of Eq.~\eqref{eq:62} by the complex conjugate
of the denominator to show that
\begin{align}
  \label{eq:63} E\sub{\Omega}=
  -2\tch\dfrac{2\sin(\dfrac{\pi}{2(L+1)})-1+\cos(\dfrac{\pi}{L+1})}{1-\cos(\dfrac{\pi}{L+1})}.
\end{align} which immediately leads to the expression
\begin{align}
  \label{eq:64}
  E\sub{\Omega}=2\tch\Bigg(1-\dfrac1{\sin\Big(\dfrac{\pi}{2(L+1)}\Big)}\Bigg).
\end{align}

Similar analyses yield the other expressions in Table~\ref{tab:2}, which compares the
ground-state energies for \obc, \pbc, and \tbc. 
In the $L\to\infty$ limit, the right-hand sides of Eqs.~\eqref{eq:57}~or \eqref{eq:58} can
be more easily computed. For \pbc, for instance, we find that
\begin{align}
  \label{eq:65} E\sub{\Omega}= \dfrac L{\pi} \int_{-\pi/2}^{\pi/2}\epsilon\sub{k}\dd k.
\end{align} Here the prefactor of the integral on the \rhs\ is the density $L/(2\pi)$ of
allowed $k$ levels in momentum space multiplied by the spin degeneracy, and the energies
$\epsilon\sub{k}$ are given by Eq.~\eqref{eq:47}. The integral on the \rhs\ of
Eq.~\eqref{eq:65} computed, we find that
\begin{align}
  \label{eq:66} E\sub{\Omega} = - \dfrac{4L}{\pi}\tch,
\end{align} which amounts to the per-particle energy in Eq.~\eqref{eq:124}.

\section{\label{sec:30} Analytical results for $U\to\infty$.}

\subsection{\label{sec:14} Open boundary condition}

Under \obc, the energy levels are given by Eq.~\eqref{eq:47}, with $k$
defined by Eq.~\eqref{eq:50}. At half filling, with $N=L$, each level is singly occupied
in the ground state. Since the distribution of energy levels is particle-hole symmetric,
the contribution of the kinetic energy to the ground-state energy vanishes, so that
\begin{align}
  \label{eq:72} E_{\Omega}^{N=L}=-\mu L.
\end{align}

By contrast, in the $N=L-1$ ground state the topmost level, with single-particle level
\begin{align}
  \label{eq:73} \epsilon\sub{k_{max}} = 2\tch\cos\Big(\frac{\pi L}{L+1}\Big),
\end{align} is vacant, and the corresponding many-body eigenvalue will include the
negative of $\epsilon\sub{k_{max}}$, that is
\begin{align}
  \label{eq:74} E^{L-1}_{\Omega}=-2\tch\cos\Big(\frac{\pi}{L+1}\Big)-\mu(L-1).
\end{align}

\subsection{\label{sec:15} Closed boundary conditions}

The exact results under \pbc\ \cite{1998IPA537,2003LiW1,2003EFG+} support the attractive
image of individual levels labeled by momenta. The same image holds under \tbc.  Either
under \pbc\ or \tbc, however, only for $U=0$ are the allowed $k$ given by
Eq.~\eqref{eq:39}.  Without Coulomb interaction, the momentum states are decoupled from
each other, and the allowed $k$ are solely determined by boundary condition. For $U\ne0$,
by contrast, the $k$ states are interdependent, and the allowed momenta depend on the spin
degrees of freedom. Even in the $U\to\infty$ limit, which is relatively simple under \obc,
as discussed in Sec.~\ref{sec:14}, the conditions determining the allowed momenta under
\pbc\ or \tbc\ depend on the total spin $S$ and its component, $S\sub{z}$.

As an illustration consider the Hamiltonian~\eqref{eq:1} with $L=4$ under \tbc\ with the
special torsion $\Theta=(\pi/2)L$, which is equivalent to \pbc, and let $U\to\infty$.  The
conservation laws divide the Fock space into sectors labeled by the charge $N$, total spin
$S$, total spin component $S\sub{z}$ and momentum $k$. We choose the chemical potential
$\mu$ so that the ground state lies in one of the sectors with $N=3$.

Coulomb repulsion forces the three electrons to occupy three distinct sites. For
definiteness, let us assume that the unoccupied state is at site $\ell=4$. The total spin
$S$ is the sum of three spin-$1/2$ variables. Each variable can have $S\sub z=\ups$ or
$S\sub z=\dos$. The three spins can therefore be found in $2^{3}=8$ configurations. The
maximum spin resulting from addition of the three variables is $S=3/2$. The minimum is
$S=1/2$. With $S=3/2$, $S\sub{z}$ can take four distinct values\textemdash a
quadruplet. Out of the eight possible configurations, four states must therefore
constitute two doublets, with $S=1/2$.

\paragraph{Quadruplet. \label{sec:33}} The $S_{z}=S=3/2$ member of the quadruplet, known as the
\emph{fully-stretched state} because the three spin components are aligned, is given by
the expression
\begin{align}
 \label{eq:75} \ket{\dfrac32,\dfrac32;\ell=4} = \cd{1\ups}\cd{2\ups}\cd{3\ups}\kvac,
\end{align} where the label $\ell=4$ on the left-hand side reminds us that the fourth site
is vacant.

Cyclic permutation of both sides of Eq.~\eqref{eq:75} yields the spin eigenstates
$\ket{3/2,3/2,\ell}$ ($\ell=1,2,3$). In analogy with Eq.~\eqref{eq:33}, we can then
construct four eigenstates of the translation operator $T\sub{1}$:
\begin{align}
  \label{eq:76} \ket{\dfrac32,\dfrac32,k} =
  \dfrac12\sum_{\ell=1}^{4}e^{-ik\ell}\ket{\dfrac32,\dfrac32;\ell}.
\end{align}

To determine the allowed momenta $k$ in Eq.~\eqref{eq:76}, we translate both sides by one
lattice parameter, that is,
\begin{align}
  \label{eq:77} T\sub{1} \ket{\dfrac32,\dfrac32,k} =
  \dfrac12\sum_{\ell=1}^{4}e^{-ik\ell}\ket{\dfrac32,\dfrac32;\ell+1},
\end{align} or if we let $\ell\to\ell-1$ in the sum on the \rhs,
\begin{align}
  \label{eq:78} T\sub{1} \ket{\dfrac32,\dfrac32,k} =
  e^{ik}\dfrac12\sum_{\ell=0}^{3}e^{-ik\ell}\ket{\dfrac32,\dfrac32;\ell}.
\end{align}

Under closed boundary condition, $\ell=0$ is equivalent to $\ell=L\equiv4$, and it follows
that
\begin{align}
  \label{eq:79} T\sub{1} \ket{\dfrac32,\dfrac32,k} = e^{ik}\ket{\dfrac32,\dfrac32,k},
\end{align} provided that $e^{ikL}=1$, which condition determines the allowed momenta:
\begin{align}
  \label{eq:80} k= -\dfrac{\pi}2,0, \dfrac{\pi}2,\pi.
\end{align}

According to Eqs.~\eqref{eq:79}~and \eqref{eq:80}, the $\ket{\dfrac32,\dfrac32,k}$ are
non-degenerate eigenstates of the translation operator $T\sub1$, which commutes with the
model Hamiltonian. It follows that the momentum eigenvectors
$\ket{\dfrac32,\dfrac32,k=n\pi/2}$ ($n=-1,\ldots,2$) are eigenstates of $\mathbf{H}$. In
fact, straightforward algebra shows that
  \begin{align}
    \label{eq:81} \mathbf{H}\ket{\dfrac32,\dfrac32,k} =
    (2\tch\sin{k}-3\mu)\ket{\dfrac32,\dfrac32,k}\qquad(k=-\dfrac{\pi}2,0,\dfrac{\pi}2,\pi).
  \end{align}

The momentum $k=-\pi/2$ yields the lowest eigenvalue,
\begin{align}
  \label{eq:82} E\sub{S=3/2}=-2\tch-3\mu.
\end{align}

Equation~\eqref{eq:81} has simple physical interpretation. The vacancy\textemdash a
hole\textemdash at site $\ell$ in the state $\ket{3/2,3/2;\ell}$ ($\ell=1,2,3,4$) can hop
to either neighboring site, $\ell-1$ or $\ell+1$, just as the electron at site $\ell$ in
Eq.~\eqref{eq:33} can hop to the neighboring sites. The spectrum of the model Hamiltonian
in the $S=S\sub{z}=3/2$ sector therefore define single-particle energies forming a band
analogous to the ones in Fig.~\ref{fig:2}(b), with single-particle energies given by
Eq.~\eqref{eq:48}.

\paragraph{Doublets.\label{sec:34}} The quadruplet~\eqref{eq:75} is unique, but the two doublets are
not. Two doublets are the symmetric combinations
\begin{align}
  \label{eq:83} \ket{\dfrac12,\dfrac12,g;\ell=4} = \dfrac1{\sqrt6}
  \Big(\cd{1\ups}\cd{2\ups}\cd{3\dos}
  -2\cd{1\ups}\cd{2\dos}\cd{3\ups}+\cd{1\dos}\cd{2\ups}\cd{3\ups}\Big),
\end{align} which is even ($g$) under left-right inversion of the lattice segment
$\ell=1,2,3$, and
\begin{align}
  \label{eq:84} \ket{\dfrac12,\dfrac12,u;\ell=4} = \dfrac1{\sqrt2}
\Big(\cd{1\ups}\cd{2\ups}\cd{3\dos}-\cd{1\dos}\cd{2\ups}\cd{3\ups}\Big),
\end{align} which is odd ($u$). To verify that the {\rhs}s are doublets, we only have to
check that $S\sub+\ket{\dfrac12,\dfrac12,p;\ell=4}=0$ ($p=g,u$), where the raising
operator is $S\sub+ \equiv \sum_{\ell}\cd{\ell\ups}\cn{\ell\dos}$.

The choices defined by Eqs.~\eqref{eq:83}~and \eqref{eq:84} are not unique, because any
linear combination between their {\rhs}s will also have spin $1/2$. One can easily verify
that they are normalized and mutually orthogonal. Cyclic permutation of
Eqs.~\eqref{eq:83}~and \eqref{eq:84} yields three other pairs with vacancies at sites
$\ell=1,2,$ and $3$, from which eight eigenstates of the translation operator $T\sub{1}$
can be constructed, as in Eq.~\eqref{eq:76}. The allowed momenta are once more given by
Eq.~\eqref{eq:80}. For each sector with $S=S\sub{z}=1/2$ and given $k$, two states
$\ket{1/2,1/2,p,k}$ ($p=g,u$) result. Projection of the model Hamiltonian upon the
orthonormal basis formed by these two states yields a $2\times2$ matrix:
\begin{align}
  \label{eq:85} \mathcal{H}\sub{S=1/2,k}=-\tch
  \begin{bmatrix}
   \sin(k)&-\sqrt3i\cos(k)\\ \sqrt3i\cos(k)&\sin(k)
  \end{bmatrix}-3\mu.
\end{align}

Diagonalization yields the two eigenvalues of the Hamiltonian in the $S=S\sub{z}=1/2$, $k$
sector:
\begin{align}
  \label{eq:86} E_{1/2,k}^{\pm}=-\tch\Big(\sin(k)\pm\sqrt3\cos(k)\Big)-3\mu.
\end{align}

The lowest eigenvalues among the four $S=S\sub{z}=1/2$, $k$ ($k=-\pi/2,0,\pi/2,\pi$)
sectors lie in the $k=0$ and $k=\pi$ sectors:
\begin{align}
  \label{eq:87} E_{1/2,0}^{+} = E_{1/2,\pi}^{-}=-\sqrt3\tch-3\mu.
\end{align}

\paragraph{Bethe-Ansatz approach. \label{sec:35}} Unfortunately, the same analysis cannot be extended to
longer lattices, because the number of basis states grows exponentially with $L$. The
alternative is the Bethe-Ansatz solution.\cite{1968LiW1445,2003LiW1}

The Bethe-Ansatz solution covers any lattice size $L$, under \obc, \pbc, or \tbc. Instead
of a closed expression for the eigenvalues of the Hamiltonian, it yields a set of coupled
nonlinear equations, known as the Lieb-Wu equations. For most choices of the model
parameters, the Lieb-Wu equations are notoriously difficult to solve, even
numerically. The exceptions are the $U=0$ limit, discussed in Section~\ref{sec:19}, the
infinite system, $L\to\infty$, to be discussed in Section~\ref{sec:21}, and the
$U\to\infty$ limit, to which we now turn.

The notation we have adopted, in which $N$ denotes the number of electrons and $M$, the
number of $\dos$-spin electrons, follows Lieb and Wu.\cite{1968LiW1445} The Bethe Ansatz
approach seeks $N$-electron eigenstates described by real-space eigenfunctions
$\Psi(x\sub{1},x\sub{2},\ldots,x\sub{N}; \sigma\sub{1},\ldots,\sigma\sub{N})$, dependent
on the particle positions $x\sub{j}$ and spin components $\sigma\sub{j}$ ($j=1,\ldots,N$).

The eigenfunctions are parametrized by two sets of quantum numbers: $k\sub{n}$
($n=1,\ldots,N$) and $\lambda\sub{m}$ ($m=1,\ldots, M$), associated with the charge and
spin degrees of freedom, respectively. To determine the $k\sub{n}$ and $\lambda\sub{m}$, a
system of $N+M$ non-linear coupled algebraic equations must be solved.

The eigenvalues of the Hamiltonian depend only on the $k\sub{n}$, which can be formally
identified with momenta. If $U=0$, the $k\sub{n}$ coincide with the single-particle
momenta $k$ in Sec.~\ref{sec:19}. With $U\ne0$ they are no longer given by
Eq.~\eqref{eq:39} or by Eq.~\eqref{eq:50} and have to be determined from the Lieb-Wu
equations.

Once the $k\sub{n}$ are found, the eigenvalues of the Hamiltonian under \obc\ or \pbc\ can
be computed from a sum analogous to Eq.~\eqref{eq:57}:\cite{1968LiW1445,2003LiW1,2003EFG+}
\begin{align}
  \label{eq:88} E= -2\tch\sum_{n=1}^{N}\cos(k\sub{n}) -\mu N,
\end{align} where the sum runs over the $N$ occupied $k\sub{n}$.

For \tbc, the sum is analogous to the \rhs\ of Eq.~\eqref{eq:58}
\begin{align}
  \label{eq:89}
    E= -2\tch\sum_{n=1}^{N}\cos(k\sub{n}+\theta) -\mu N,
\end{align} which for the special torsion $\Theta\equiv L\theta=(\pi/2)L$ reads
\begin{align}
  \label{eq:90}
    E= 2\tch\sum_{n=1}^{N}\sin(k\sub{n}) -\mu N.
\end{align}

The chemical potential is determined by the condition $\partial \bar E/\partial N=0$,
where $\bar E$ is the thermodynamical average of the eigenvalues $E$. At zero temperature,
$\mu$ is such that the $N$ occupied $k\sub{n}$ satisfy the inequality
$-2t\cos(k\sub{n})\le\mu$ ($n=1,\ldots, N$).

\begin{figure}[h!]
  \centering
  \includegraphics[width=\columnwidth]{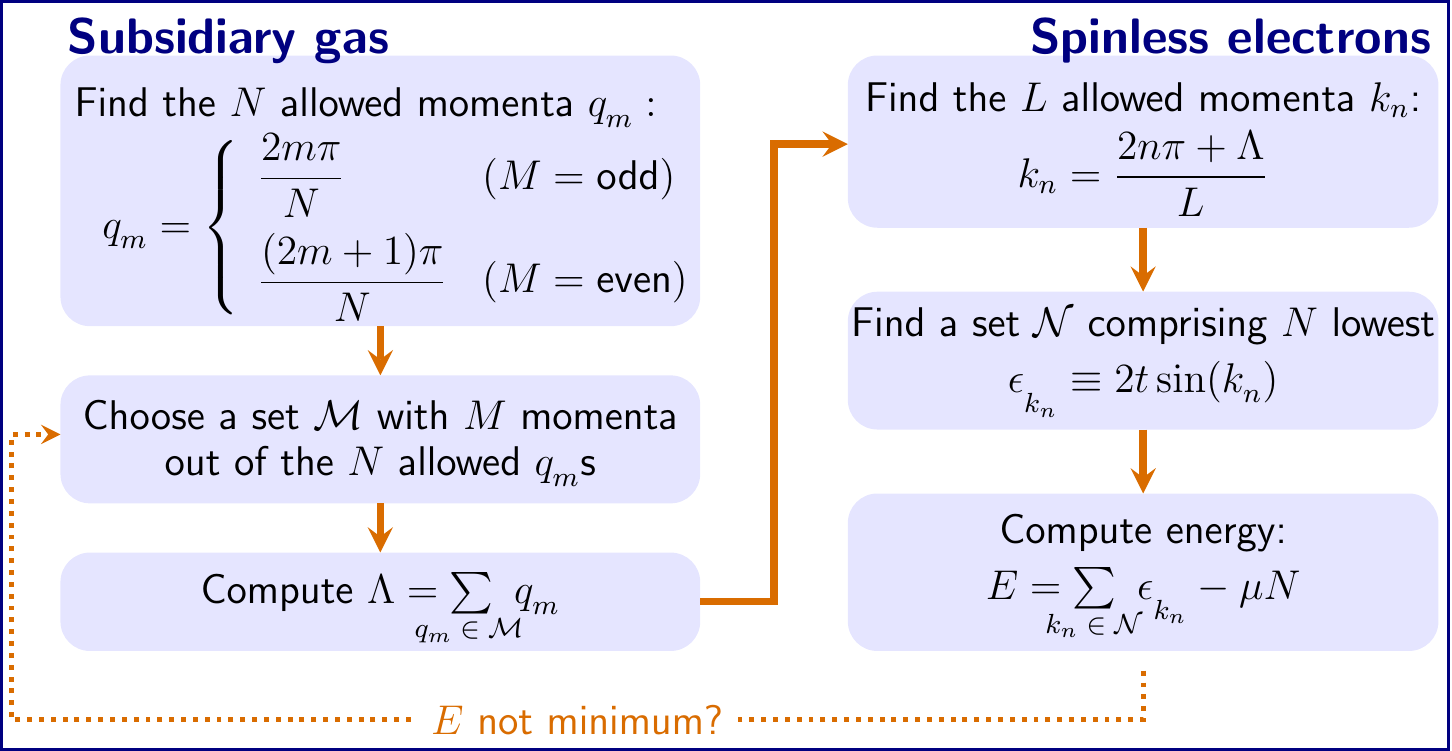}
  \caption[\ ]{Computation of the ground-state energy from the solution of the Lieb-Wu equations in
    the $U\to\infty$ limit, under \tbc\ with the special torsion $\Theta=(\pi/2)L$. $L$, $N$, and
    $M$ are the lattice size, the number of electrons, and the number of $\dos$-electrons
    respectively. The ground-state energy is computed from Eq.~\eqref{eq:90}, where the $k\sub{n}$,
    given by Eq.~\eqref{eq:92}, can be regarded as momenta of spinless electrons. To determine the
    phase $\Lambda$, one starts out by considering a subsidiary gas of non-interacting particles
    with momenta $q\sub{m}$. The $M$ integers $m$ are chosen so that the resulting $q\sub{m}$, given
    by Eq.~\eqref{eq:93}, lie in the First Brillouin Zone. Given the $q\sub{m}$, Eq.~\eqref{eq:95}
    determines the phase $\Lambda$. The next steps are depicted on the right-hand panel. We start by
    determining the $L$ allowed momenta $k\sub{n}$. The integers $n$ are chosen to position the
    $k\sub{n}$ in the First Brillouin Zone and to minimize the energy in Eq.~\eqref{eq:89}. The
    resulting minimum energy $E\sub{\mathcal{M}}$ depends on $\Lambda$ and hence upon our choice of the set
    $\mathcal{M}$. To find the ground-state energy, we have to repeat the procedure for all possible
    $\mathcal{M}$s. The lowest overal $E\sub{\mathcal{M}}$ is the ground-state energy.}
  \label{fig:4}
\end{figure} The $U\to\infty$ limit simplifies the Lieb-Wu equations. A schematic
depiction of the procedure determining the ground-state energy is presented in
Fig.~\ref{fig:4}. The charge and spin degrees of freedom decouple and can be described
separately. The $k\sub{n}$ satisfy a relatively simple equation, analogous to
Eq.~\eqref{eq:36}:\cite{1968LiW1445,2003LiW1,2003EFG+}
\begin{align}
  \label{eq:91} e^{ik\sub{n}L}=e^{i\Lambda},
\end{align} where the phase $\Lambda$ depends only on the spin degrees of freedom.

Equation~\eqref{eq:91} allows momenta of the form
\begin{align}
  \label{eq:92} k\sub{n} = \dfrac{2\pi n+\Lambda}{L},
\end{align} where the $n$'s are integers that define the eigenstate of the
Hamiltonian. The integers defining the ground state for \tbc, for example, are those that
minimize the sum on the \rhs\ of Eq.~\eqref{eq:90}.

To determine the allowed momenta, we therefore need the phase $\Lambda$ and have to
examine the spin degrees of freedom. Again, we let the number $M$ of electrons with $\dos$
spin be smaller or equal to the number $N-M$ of $\ups$ electrons. Although the Lieb-Wu
equation describing the spin degrees of freedom seem unwieldy, they have been found to be
identical with the equations describing a simpler system, a subsidiary gas with a
Hamiltonian that can be trivially diagonalized.\cite{1998IPA537} The eigenvalues of the
latter Hamiltonian determine the phase $\Lambda$, which can then be substituted on the
\rhs\ of Eq.~\eqref{eq:92} to yield the allowed momenta $k\sub{n}$.

More specifically, to determine $\Lambda$ one has to find the total momentum of a subsidiary system
with $M$ particles on an $N$-site one-dimensional lattice. The particles in the subsidiary
system occupy $M$ distinct states labeled by their momenta $q\sub{m}$ (where $1\le m\le N$), which
lie on a flat band, with dispersion relation $\epsilon\sub{q}=0$. The subsidiary particles must
satisfy either anti-periodic or periodic boundary conditions, depending on whether $M$ is even or
odd, respectively. The $M$ allowed momenta must therefore satisfy the equalities
\begin{align}
  \label{eq:93} e^{iq\sub{m}N}=
  \begin{cases}
    -1&\qquad(M=\mbox{even})\\
    1 &\qquad(M=\mbox{odd}),
  \end{cases}
\end{align} which are equivalent to the expressions
\begin{align}
  \label{eq:94} q\sub{m} =
  \begin{cases}
    \dfrac{(2m+1)\pi}{N}&\qquad(M=\mbox{even})\\[2mm]
    \dfrac{2m\pi}{N}&\qquad(M=\mbox{odd})
  \end{cases},
\end{align} with integers $1\le m\le N$ that depend on the desired eigenstate of the Hamiltonian.

Given a set of $M$ occupied momenta $q\sub{m}$, the phase $\Lambda$ is the total momentum
\begin{align}
  \label{eq:95} \Lambda = \sum_{m=1}^{M}q\sub{m}.
\end{align}

This explained, we are ready to find the eigenvalues of the $L=4$, $U\to\infty$ Hubbard
Hamiltonian for $N=3$.

First, we set $M=0$, which is equivalent to letting $S\sub{z}=S= 3/2$. With $M=0$, the
number of particles in the subsidiary gas is zero and it follows from Eq.~\eqref{eq:95}
that $\Lambda=0$.  As in Sec.~\ref{sec:11}, we choose the $k\sub{n}$ to lie in the first
Brillouin Zone. Equation~\eqref{eq:92} then yields the allowed momenta:
\begin{align}
  \label{eq:96} k\sub{n} = \dfrac{\pi n}2\qquad(n=-1,0,1,2).
\end{align}

To obtain the smallest eigenvalue of the Hamiltonian associated to the $k\sub{n}$ in
Eq.~\eqref{eq:96}, we fill the three levels making the smallest contribution to the \rhs\
of Eq.~\eqref{eq:90}, i.~e., the levels associated with $k\sub{-1}, k\sub{0}$ and
$k\sub{2}$. The resulting eigenvalue coincides with the \rhs\ of Eq.~\eqref{eq:82}.

Consider now $M=1$. With $M=1$, the $q\sub{m}$ allowed by Eq.~\eqref{eq:94} are
\begin{align}
  \label{eq:97} q\sub{m} = \dfrac{2m\pi}3\qquad(n=-1,0,1).
\end{align}

Equation~\eqref{eq:95} then determines $\Lambda$. Since $M=1$, the sum on the \rhs\ is
restricted to a single $q\sub{m}$, namely one of the three values in
Eq.~\eqref{eq:97}. The resulting phases are given by the equality
\begin{align}
  \label{eq:98} \Lambda = -\dfrac{2\pi}3, 0, \dfrac{2\pi}3.
\end{align}

Substitution of the \rhs\ of Eq.~\eqref{eq:98} for $\Lambda$ in Eq.~\eqref{eq:92} yields
the following allowed momenta:
\begin{align}
  \label{eq:99} k =
  \begin{cases}
    -\dfrac{2\pi}3, -\dfrac{\pi}6, \dfrac{\pi}3,
    -\dfrac{5\pi}6&\qquad(\Lambda=-\dfrac{2\pi}3)\\[3mm] \dfrac{\pi}2, 0, \dfrac{\pi}2,
    -\pi&\qquad(\Lambda=0)\\[3mm] \dfrac{\pi}3, -\dfrac{5\pi}6, \dfrac{\pi}6,
    -\dfrac{2\pi}3&\qquad(\Lambda=\dfrac{2\pi}3)
  \end{cases}.
\end{align}

To obtain the corresponding eigenvalues, from Eq.~\eqref{eq:90}, for each $\Lambda$ we
have to occupy three of the four allowed $k$-states, i.~e., leave one level vacant. The
resulting energies are given by the equality
\begin{align}
  \label{eq:100} E+3\mu =
  \begin{cases}
    \pm\sqrt3\tch, \pm \tch&\qquad(\Lambda=\pm\dfrac{2\pi}3)\\ 0, -2\tch, 2\tch
    &\qquad(\Lambda=0),
  \end{cases}
\end{align} the eigenvalues for $\Lambda=2\pi/3$ being degenerate with those for
$\Lambda=-2\pi/3$, and the first eigenvalue for $\Lambda=0$ being doubly degenerate. The
lowest eigenvalues for $\Lambda=\pm2\pi/3$ and for $\Lambda=0$ are $-\sqrt3\tch-3\mu$ and
$-2\tch-3\mu$, respectively.

Comparison of Eq.~\eqref{eq:100} with Eqs.~\eqref{eq:82}~and \eqref{eq:87} shows that with
$M=1$ the phase $\Lambda=0$ corresponds to $S=3/2$, $S\sub{z}=1/2$ [Eq.~\eqref{eq:82}],
while $\Lambda=\pm2\pi/3$ corresponds to $S=S\sub{z}=1/2$ [Eq.~\eqref{eq:87}]. This
concludes our illustrative discussion.

The same procedure can be applied to other lattice lengths $L$ and electron numbers
$N$. We are especially interested in the minimum energies in the sectors with $N=L$ and
$N=L-1$, from which we can compute the $U\to\infty$ ground-state energy $E\sub{\Omega}$
and the energy gap $E\sub{g}$ at half filling.

With $N=L$, the ground-state energy vanishes in the $U\to\infty$ limit.  Since each
$k$-level can host at most one electron, all levels must be occupied for
$N=L$. Particle-hole symmetry then guarantees that the positive contributions to
$E\sub{\Omega}$ cancel the negative contributions. The ground-state energy is therefore
zero.

With $N=L-1$, except for the special length $L=2$, the ground-state energy is
negative. For fixed $\Lambda$, Eq.~\eqref{eq:92} defines the allowed momenta. In the
ground state all levels are filled, except for the highest one, with energy
$\epsilon\sub{max}$. The ground-state energy is $-\epsilon\sub{max}$.  With $\Theta=(\pi
L)/2$, provided that the momentum $k\sub{n}=\pi/2$ be allowed, the highest allowed energy
is $\epsilon\sub{max}=\epsilon\sub{k\sub{n}=\pi/2}=2\tch$.  If $k\sub{n}=\pi/2$ is not
allowed, the ground-state energy will be $-2\tch\sin(\bar k)$, where $\bar k$ is the
allowed momentum closest to $\pi/2$.

For lengths $L$ that are multiples of four, one of the momenta allowed by
Eq.~\eqref{eq:92} is $k\sub{n}=\pi/2+\Lambda/L$. The phase $\Lambda=0$ is always allowed,
since we can always choose $M=0$. The momentum $k\sub{n}=\pi/2$ is therefore allowed, and
the ground-state energy is $-2\tch$.

The ground-state energy is also $-2\tch$ if $N=L-1$ is a multiple of four. Given
$\Lambda$, the momentum $k\sub{n=0}=\Lambda$ is always allowed by Eq.~\eqref{eq:92}. We
choose $M=1$. According to Eq.~\eqref{eq:94}, the subsidiary momentum $q\sub{N/4}=\pi/2$
is allowed, and hence the phase can take the value $\Lambda=\pi/2$. It follows that
$k\sub{n}=\pi/2$ is allowed, and that the ground-state energy is $-2\tch$.

If neither $L$ nor $N$ are multiples of four, $k\sub{n}$ cannot equal $\pi/2$, and the
ground-state energy $E\sub{\Omega}$ is positive. To compute it we must first let $M$ run
from zero to $N$, consider all subsidiary momenta $q\sub{m}$ momenta compatible with
Eq.~\eqref{eq:94} for each $M$ and obtain the resulting phases $\Lambda$ from
Eq.~\eqref{eq:95}. Once the $\Lambda$ are computed, the allowed $k\sub{n}$ are given by
Eq.~\eqref{eq:92}. The ground-state energy under \tbc\ is given by the set of $N$ momenta
$k\sub{n}$ thus determined that minimizes the right-hand side of Eq.~\eqref{eq:90}.

\subsection{Ground-state energy for $L\to\infty$}
\label{sec:21}

As $L\to\infty$, the quantum numbers $k\sub{n}$ and $\lambda\sub{m}$ characterizing the
Bethe-Ansatz solution form continua. When the ground state is considered, the Lieb-Wu
equations reduce to two coupled integral equations for the densities of the $k\sub{n}$ and
$\lambda\sub{n}$. For the special case $2M=N=L$, i.~e., for the spin-unpolarized
half-filled band, Lieb and Wu were able to solve the integral equations and derive closed
expressions for the ground-state energy $E\sub{\Omega}$ and chemical
potential.\cite{1968LiW1445,2003LiW1,2003EFG+} Their expression for the ground-state
energy, which excludes the contribution from the term proportional to $\mu$ on the \rhs\
of Eq.~\eqref{eq:1}, reads
\begin{align}
  \label{eq:101} E^{LW}_{\Omega,N=L} =
  -4L\int_{0}^{\infty}
  \dfrac{J\sub{0}(\omega)J\sub{1}(\omega)}{\omega\Big(1+e^{\omega U/2}\Big)} \dd\omega
\end{align} where $J\sub{\nu}$ denotes the $\nu$-th order Bessel function.

The chemical potential, defined as the energy difference
$E^{LW}_{\Omega,N+1}-E^{LW}_{\Omega,N}$ needed to add a particle to the ground state, is
given by the equality
\begin{align}
  \label{eq:102} \mu\sub{+} = \dfrac{U}2 - 2 +
  4\int_{0}^{\infty}\dfrac{J\sub{1}(\omega)}{\omega\Big(1+e^{\omega U/2}\Big)}\dd\omega.
\end{align}

\subsection{Energy gap for $L\to\infty$}
\label{sec:23}

The subscript $+$ on the \lhs\ of Eq.~\eqref{eq:102} is necessary, because the chemical
potential is discontinuous for $U\ne0$. The chemical potential $\mu\sub{-}$, equal to the
energy $E^{LW}_{\Omega,N}-E^{LW}_{\Omega,N-1}$ needed to add a particle to the
$N-1$-electron ground state, can be obtained from the particle-hole transformation in
Sec.~\ref{sec:8}:
\begin{align}
  \label{eq:103} \mu\sub{-} = U -\mu\sub{+}.
\end{align}

The energy gap $E\sub{g}=\mu\sub{+}-\mu\sub{-}$ is therefore given by the closed
expression
\begin{align}
  \label{eq:104}
    E\sub{g} = U - 4 + 8\int_{0}^{\infty}\dfrac{J\sub{1}(\omega)}{\omega\Big(1+e^{\omega
    U/2}\Big)}\dd\omega,
\end{align} the \rhs\ of which vanishes as $U\to0$.

\bibliography{bjp}
\end{document}